\documentclass[11pt,a4paper]{article}
\usepackage{authblk}
\usepackage{lmodern}
\usepackage{comment}
\usepackage{jheppub}
\usepackage{amsmath}
\usepackage{braket}
\usepackage{mathtools}
\usepackage{tikz}
\usepackage{empheq}
\usepackage{enumitem}
\usepackage{graphics}
\usepackage{wrapfig}
\usepackage{caption}
\usepackage{natbib}
\usepackage{xcolor}
\usepackage{framed}
\usepackage{array}
\usepackage{dsfont}
\definecolor{shadecolor}{gray}{0.925}

\usepackage[cbgreek]{textgreek}

\def\sideremark#1{\ifvmode\leavevmode\fi\vadjust{\vbox to0pt{\vss
 \hbox to 0pt{\hskip\hsize\hskip1em
 \vbox{\hsize3cm\tiny\raggedright\pretolerance10000
 \noindent #1\hfill}\hss}\vbox to8pt{\vfil}\vss}}}%

\newcommand{\bi}{\begin{itemize}}
\newcommand{\ei}{\end{itemize}}
\newcommand{\bea}{\begin{align}}
\newcommand{\eea}{\end{align}}
\newcommand{\be}{\begin{equation}}
\newcommand{\ee}{\end{equation}}

\makeatletter
\renewcommand*\env@matrix[1][\arraystretch]{%
  \edef\arraystretch{#1}%
  \hskip -\arraycolsep
  \let\@ifnextchar\new@ifnextchar
  \array{*\c@MaxMatrixCols c}}
\makeatother
\author[\ensuremath{a}]{Alistair J. CHOPPING}
\author[\ensuremath{a},\ensuremath{b},\ensuremath{c}]{\quad Charlotte SLEIGHT}
\author[\ensuremath{b},\ensuremath{c},\ensuremath{d}]{\quad Massimo TARONNA}

\affiliation[\ensuremath{a}]{Centre for Particle Theory and Department of Mathematical Sciences, \\ Durham University, Durham, DH1 3LE, U.K.}

\affiliation[\ensuremath{b}]{Dipartimento di Fisica ``Ettore Pancini'', Universit\`a degli Studi di Napoli Federico II, \\Monte S. Angelo, Via Cintia, 80126 Napoli, Italy}

\affiliation[\ensuremath{c}]{INFN, Sezione di Napoli, Monte S. Angelo, Via Cintia, 80126 Napoli, Italy}

\affiliation[\ensuremath{d}]{Scuola Superiore Meridionale, Universit\`a degli Studi di Napoli Federico II,\\ Largo San Marcellino 10, 80138 Napoli, Italy}

\emailAdd{alistair.j.chopping@durham.ac.uk,charlotte.sleight@na.infn.it, massimo.taronna@unina.it}

\title{\centering \Huge Cosmological Correlators \\ for Bogoliubov Initial States}

\abstract{We consider late-time correlators in de Sitter (dS) space for initial states related to the Bunch-Davies vacuum by a Bogoliubov transformation. We propose to study such late-time correlators by reformulating them in the familiar language of Witten diagrams in Euclidean anti-de Sitter space (EAdS), showing that they can be perturbatively re-cast in terms of corresponding dS boundary correlators in the Bunch-Davies vacuum and in turn, Witten diagrams in EAdS. Unlike the standard relationship between late-time correlators in the Bunch-Davies vacuum and EAdS Witten diagrams, this involves points on the upper and lower sheet of the EAdS hyperboloid which account for antipodal singularities of the two-point functions. Such Bogoliubov states include an infinite one parameter family of de Sitter invariant vacua as a special case, where the late-time correlators are constrained by Conformal Ward identities. In momentum space, it is well known that their late-time correlators exhibit singularities in collinear (``folded") momentum configurations. We give a position space interpretation of such solutions to the conformal Ward identities, where in embedding space they can be generated from the solution without collinear singularities by application of the antipodal map. We also discuss the operator product expansion (OPE) limit of late-time correlators in a generic dS invariant vacuum. Many results are derived using the Mellin space representation of late-time correlators, which in this work we extend to accommodate generic dS invariant vacua.}

\begin{document}

\begin{flushright}    
\texttt{}
\end{flushright}

\maketitle

\newpage

\section{Introduction}

De Sitter (dS) space is a maximally symmetric space-time of constant positive curvature and provides one of the simplest models of an exponentially expanding universe. There is observational evidence for two periods of exponential expansion; the current accelerated expansion and the period of inflation moments after the big bang. The latter would be the earliest and highest energy observable process in the history of our universe, where measurements of spatial correlations between cosmological structures today can be traced back to the end of inflation - on the future boundary of an approximate dS space.

\vskip 4pt
The standard approach to make predictions for inflationary correlation functions is to use the ``in-in" formalism in QFT \cite{Maldacena:2002vr,Weinberg:2005vy}, which has given rise to the first perturbative results \cite{Creminelli:2003iq,Seery:2005wm,Chen:2006nt,Seery:2006vu,Seery:2008ax,Chen:2009zp,Adshead:2009cb} for late time correlators and a rich spectrum of inflationary phenomenology. In recent years the ``Cosmological Bootstrap"\footnote{For an overview of Cosmological Bootstrap programme see e.g. \cite{Baumann:2022jpr}.} has emerged as a complementary approach which aims to construct such inflationary correlators directly on the future boundary using physical principles such as unitarity, locality and symmetry as consistency requirements. Invariance under the full set of dS isometries in particular implies that correlation functions on the late-time boundary of de Sitter space are constrained by conformal Ward identities \cite{Antoniadis:2011ib,Maldacena:2011nz,Creminelli:2011mw,Bzowski:2011ab,Kehagias:2012pd,Kehagias:2012td,Schalm:2012pi,Bzowski:2012ih,Mata:2012bx,Bzowski:2013sza,Ghosh:2014kba,Kundu:2014gxa,Arkani-Hamed:2015bza,Shukla:2016bnu,Arkani-Hamed:2018kmz}. The latter, combined with consistent on-shell factorisation and choice of initial state, has led to the first complete analytic understanding of the four-point function for massless scalars mediated by the exchange of a massive scalar in the Bunch-Davies vacuum \cite{Arkani-Hamed:2018kmz}. 

\vskip 4pt
De Sitter space in fact admits an infinite one-parameter family of vacuum states that are all invariant under the de Sitter isometries \cite{Allen:1985ux,Mottola:1984ar}, which are often referred to  as ``$\alpha$-vacua" with superselection parameter $\alpha$. The Bunch-Davies vacuum, corresponding to $\alpha=0$, is the unique vacuum that extrapolates to the standard Minkowski vacuum in the limit where the cosmological constant vanishes. Two-point functions in the Bunch-Davies vacuum therefore have the usual short-distance singularity along the light cone, while the other dS invariant vacua (also) have a singularity when the points are antipodal to one another. The latter is rather unconventional since in dS antipodal points are separated by a cosmological horizon and whether or not interacting QFTs in vacua with $\alpha \ne 0$ are consistent has been the subject of debate \cite{Danielsson:2002mb,Banks:2002nv,Einhorn:2002nu,Kaloper:2002cs,Goldstein:2003ut,Einhorn:2003xb,Collins:2003zv,Goldstein:2003qf}. For these reasons, late-time correlators with $\alpha \ne 0$ are less studied than their Bunch-Davies counterparts -- though their phenomenological significance has been discussed e.g. in \cite{Danielsson:2002kx,Danielsson:2002qh,Goldstein:2002fc,Xue:2008mk,Ashoorioon:2013eia,Ashoorioon:2018sqb,Kanno:2022mkx,Gong:2023kpe}. From a holographic perspective, it has been argued that in dS/CFT the bulk $\alpha$-vacua correspond to a one-parameter family of marginal deformations of the putative dual CFT on the boundary \cite{Bousso:2001mw,Spradlin:2001nb}. 

\vskip 4pt
From the perspective of the Cosmological Bootstrap, most studies of late-time correlators in $\alpha$-vacua have focused on symmetry considerations; in \cite{Shukla:2016bnu} it was confirmed in some examples that their late-time correlators in de Sitter space satisfy the momentum space conformal Ward identities (as expected). In inflation the situation is more complicated, where scale invariance is broken by the time dependence of the background geometry. In slow roll inflation however the symmetry breaking is controlled and conformal symmetry is still constraining \cite{Mata:2012bx,Ghosh:2014kba,Kundu:2014gxa,Kundu:2015xta}. For scale invariance, this gives rise to the celebrated Maldacena consistency relation \cite{Maldacena:2002vr,Creminelli:2004yq}. Interestingly, for inflationary correlators in $\alpha$-vacua it has been observed \cite{Shukla:2016bnu} that the Maldacena consistency condition does not seem to be satisfied by three point functions of scalar perturbations for $\alpha \ne 0$, though positive results have been obtained very recently at four-points \cite{Ansari:2024pgq}. Other works within the scope of the Cosmological Bootstrap include a cosmological analogue of Cutkosky's cutting rules \cite{Goodhew:2020hob,Melville:2021lst}, deriving from unitarity, which have been extended to $\alpha$-vacua as well \cite{Cespedes:2020xqq,Ghosh:2024aqd}.

\vskip 4pt
Another approach to the Cosmological Bootstrap has been to try to draw lessons from the relatively well understood anti-de Sitter (AdS) case, where boundary correlation functions are, through the AdS/CFT correspondence, constrained non-perturbatively by the axioms of the Conformal Bootstrap \cite{Poland:2018epd,Kravchuk:2021kwe}. Using the structural similarities between dS and AdS, perturbative late-time correlators in the in-in formalism can be reformulated in the more familiar language of Witten diagrams in Euclidean Anti-de Sitter (EAdS) space. It is well known that dS and EAdS are related under analytic continuation, which in turn implies that dS propagators are a linear combination of appropriately analytically continued solutions to the propagator equation for a field of the same quantum numbers in EAdS. These relations have been understood for late-time correlators in the Bunch-Davies vacuum \cite{Sleight:2019hfp,Sleight:2020obc,Sleight:2021plv}, and from them it follows that any given perturbative contribution can be expressed in terms of corresponding Witten diagrams in EAdS. Late-time correlators in the Bunch-Davies vacuum therefore have a similar analytic structure to their AdS counterparts in the Euclidean regime, and this has led to various new insights \cite{Sleight:2020obc,Sleight:2021plv,Hogervorst:2021uvp,DiPietro:2021sjt,Schaub:2023scu,Loparco:2023rug,Loparco:2023akg,Chowdhury:2023arc,DiPietro:2023inn} on their properties - including at the non-perturbative level. This also opens up the possibility to import techniques and results from the relatively well-understood AdS case.

\vskip 4pt
Recasting late-time correlators in a generic $\alpha$-vacuum into the more familiar language of EAdS Witten diagrams may therefore lead to new insights on their structure, as well as new techniques to evaluate them. To achieve this, we note that propagators in the in-in formalism for generic $\alpha$ can be expressed in terms of their counterparts in the Bunch-Davies vacuum using the antipodal transformation, building on a similar observation \cite{Allen:1985ux} made in the early literature. We use this to show, to all orders in perturbation theory, that late-time correlators in a generic de Sitter invariant vacuum can be expressed as a linear combination of corresponding contributions to late-time correlators in the Bunch-Davies vacuum with points antipodally transformed. In momentum space the antipodal transformation corresponds to flipping the magnitudes of the boundary momenta by phases $e^{\pm \pi i}$. Using the known relationship \cite{Sleight:2019hfp,Sleight:2020obc,Sleight:2021plv} between perturbative Bunch-Davies correlators and Witten diagrams in EAdS, this gives rise to an EAdS reformulation of late-time correlators in a generic de Sitter invariant vacuum, where the antipodal transformation maps points from the upper to the lower sheet of the EAdS hyperboloid. These results moreover suggest a position space interpretation of the solutions to conformal Ward identities in momentum space that have singularities for collinear momentum configurations, how they arise from a bulk perspective and why they are discarded in the context of the AdS/CFT correspondence.

\vskip 4pt
The family of $\alpha$-vacua are in fact a special (de Sitter invariant) case of a more general class of states that are Bogoliubov transformations of the Bunch-Davies $\left(\alpha=0\right)$ state. These are known in the literature as $\alpha$-states \cite{Einhorn:2003xb,deBoer:2004nd} or Bogoliubov states, where the parameter $\alpha$ can depend on the momentum and dS invariance is broken. This momentum dependence can be chosen so that such states are excited (squeezed) states on top of the Bunch-Davies vacuum. In momentum space our results straightforwardly extend to these more general Bogoliubov initial states simply by replacing $\alpha \to \alpha_k$ for each mode of momentum $k$. The phenomenological significance of late-time correlators for such initial states has been discussed in \cite{Starobinsky:2001kn,Starobinsky:2002rp,Easther:2001fz,Easther:2002xe,Brandenberger:2002hs,Holman:2007na,Meerburg:2009ys,Agullo:2010ws,Ganc:2011dy,Kundu:2013gha,Akama:2020jko,Ghosh:2022cny,Akama:2023jsb}.

\vskip 4pt
We apply the above relations with EAdS Witten diagrams in various examples, focusing for simplicity on non-derivative interactions. We obtain novel results for late-time correlators in $\alpha$-vacua for scalar fields of generic mass in dS$_{d+1}$, as well as making contact with some existing results for massless fields in $d=3$. We study the operator product expansion (OPE) limit of three-point functions where, for a generic $\alpha$-vacuum, the contribution corresponding to a massive particle appears to be enhanced with respect to the Bunch-Davies ($\alpha=0$) case. We argue that $\alpha$-vacua correlators are consistent with the OPE limit, in contrast to previous reports in the literature.

\vskip 4pt
Many of the results presented in this work are derived using the Mellin space representation \cite{Sleight:2019mgd,Sleight:2019hfp,Sleight:2020obc,Sleight:2021plv}, which we extend to late-time correlators in a generic de Sitter invariant vacuum in this work.

\vskip 4pt
The main results of this work are the following:

\begin{itemize}
   \item {\bf Correlators for Bogoliubov initial states from Bunch-Davies} (section \ref{subsec::alphafromBD}). Perturbative contributions to late-time correlators for a generic $\alpha$-vacuum in dS$_{d+1}$ can be re-written in terms of their counterparts in the Bunch-Davies vacuum with points antipodally transformed to the boundary at past infinity. In momentum space this corresponds to flipping the magnitudes of the boundary momenta by phases $e^{\pm \pi i}$. This is proven for massive scalar fields (i.e. principal series representations) and we show that it is possible to obtain expressions for other representations by analytic continuation and, where appropriate, renormalisation. Upon replacing $\alpha \to \alpha_k$ for each mode of momentum $k$, this extends to more general Bogoliubov initial states.

    \item {\bf Correlators for Bogoliubov initial states from EAdS Witten diagrams} \newline (section \ref{subsec::toEAdS}). Perturbative late-time correlators for a generic $\alpha$-vacuum in dS$_{d+1}$ can be reformulated in terms of corresponding Witten diagrams in EAdS$_{d+1}$ with points antipodally transformed from the boundary on the upper to the lower sheet of the EAdS hyperboloid. As above, in momentum space this corresponds to flipping the magnitudes of the boundary momenta by phases $e^{\pm \pi i}$ and extends to more general Bogoliubov initial states upon replacing $\alpha \to \alpha_k$ for each mode of momentum $k$. This extends the results \cite{Sleight:2019hfp,Sleight:2020obc,Sleight:2021plv} for late-time correlators in the Bunch-Davies vacuum to initial states related to the latter via Bogoliubov transformation. Various examples for contact and exchange diagrams in $\alpha$-vacua are given in section \ref{sec::PC}.

     \item {\bf OPE limit of correlators in generic dS invariant vacua}  (section \ref{sec::PC}). We study the operator product expansion (OPE) limit of three-point functions in a generic $\alpha$-vacuum, where the contribution corresponding to a massive particle appears to be enhanced with respect to the Bunch-Davies ($\alpha=0$) case. In contrast to previous reports in the literature, we argue that $\alpha$-vacua correlators are consistent with the OPE limit.
    
    \item {\bf Solutions to conformal Ward identities} (section \ref{sec::CWI}). We give a position space interpretation of solutions to conformal Ward identities with collinear singularities in momentum space. This is particularly clear-cut in the embedding space formulation \cite{Costa:2011mg} of CFT, where solutions to the conformal Ward identities that are singular in collinear momentum configurations correspond to the antipodal transformation of one or more of the points on the projective null cone.

    \item {\bf Mellin space for generic dS invariant vacua} (appendix \ref{app::Mellin}). We extend the Mellin space representation of momentum space conformal correlators \cite{Sleight:2019mgd,Sleight:2019hfp,Sleight:2020obc,Sleight:2021plv} to include late-time correlators in a generic $\alpha$-vacuum, from both a bulk and boundary perspective. These differ from their Bunch-Davies ($\alpha=0$) counterparts by phases in the Mellin variables. We show how to solve the momentum space conformal Ward identities using the Mellin representation, where dilatation symmetry is manifest and the Ward identity for special conformal transformations is reduced to a recursion relation. For this reason, various propagator and correlator identities appearing in this work are manifest in Mellin space.
\end{itemize}

\paragraph{Outline.} The paper is organised as follows. We begin in section \ref{sec::dSST} with an overview of the geometry of de Sitter space, focusing on two charts in particular; the Expanding Poincar\'e Patch (EPP) and the Contracting Poincar\'e Patch (CPP). In section \ref{sec::SFEPP} we review the salient aspects of scalar field theory in the EPP, and detail how the $\alpha$-vacua arise. We compute the bulk-to-bulk and bulk-to-boundary propagators in a generic Bogoliubov initial state, and show how these can be written in terms of their counterparts in the Bunch-Davies vacuum. Section \ref{sec::LTCF} begins with a short review of the in-in formalism for computing late-time correlation functions in de Sitter space before showing that, to all orders in perturbation theory, late-time correlators involving scalar fields for a generic Bogoliubov initial state can be expressed in terms of their Bunch-Davies counterparts with flips in the signs of the magnitudes of the boundary momenta. Building on previous work \cite{Sleight:2019hfp,Sleight:2020obc,Sleight:2021plv}, we then derive Feynman rules for recasting the dS propagators for late-time correlators with a generic choice of Bogoliubov initial state in terms of propagators for corresponding boundary correlators in Euclidean Anti-de Sitter, with flips in the magnitudes of the boundary momenta. In section \ref{sec::PC} we apply the Feynman rules from sections \ref{sec::SFEPP} and \ref{sec::LTCF} to re-write perturbative contributions to late-time $\alpha$ vacuum correlators in terms of corresponding Witten diagrams in EAdS in various examples. We also study the OPE limit of three-point functions in $\alpha$-vacua. In section \ref{sec::CWI} we give a position space interpretation of the solutions to conformal Ward identities that have singularities for collinear momentum configurations. In particular we show that such solutions can be generated from the solution with no collinear singularity by application of the antipodal map. 

Technical details regarding various propagator and correlator identities, as well as comments on regularisation are left to the appendices.
\subsection{Notations and Conventions}
We work in $(d+1)$-dimensional (EA)dS with the ``mostly plus'' signature. Momentum vectors are denoted $\vec{k}$, and have magnitude $k\equiv\vert\vec{k}\vert$. The momentum of the $i$-th external leg in a correlation function is denoted $\vec{k}_i$, while $\vec{k}$ is reserved for the momentum of an exchanged particle. Ambient space indices are denoted by capital Latin letters $A, B\in\{0,1,...,d+1\}$. Bulk scalar fields are denoted by $\phi$, and have scaling dimension $\Delta=\frac{d}{2}+i\nu$ unless otherwise stated. In the appendix, $s$ and $u$ are reserved for Mellin variables associated with external and internal particles, respectively. Unless otherwise stated we set the curvature radius $L$ of (EA)dS to $L=1$.

\newpage

\section{de Sitter space-time}
\label{sec::dSST}

\begin{figure}[htb]
    \centering
    \includegraphics[width=0.6\textwidth]{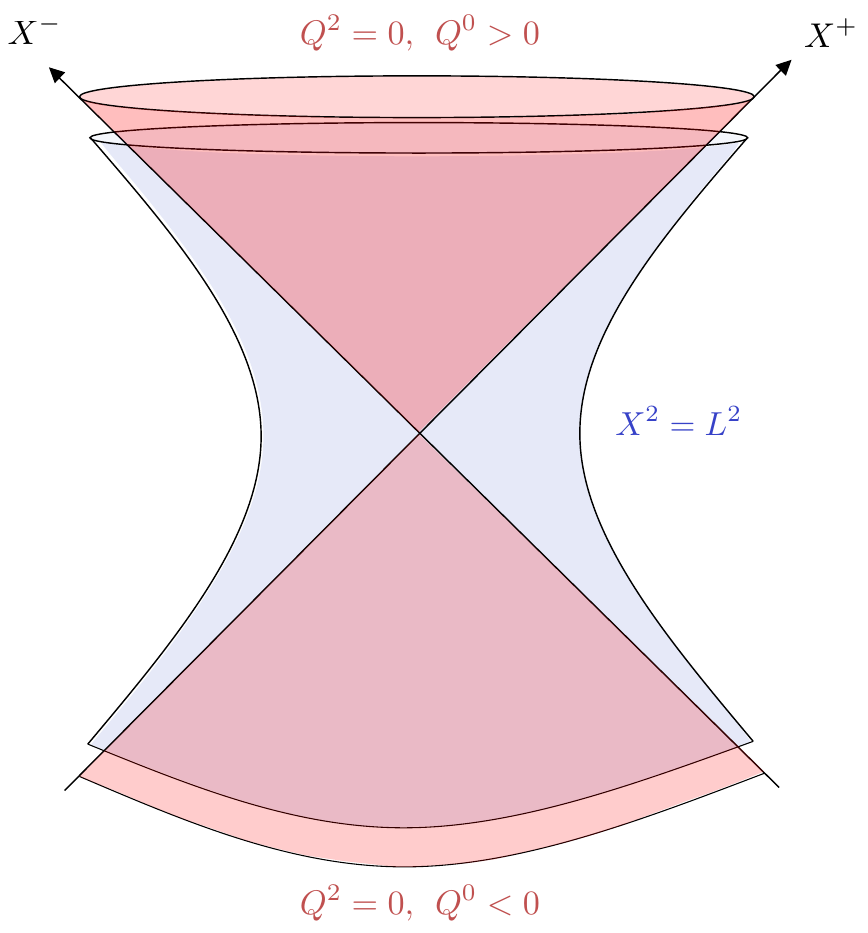}
    \caption{dS space $X^2=L^2$ with curvature radius $L$ and its boundaries $Q^2=0$, $Q \equiv \lambda Q$, $\lambda \in \mathbb{R}^+$, at past infinity $(Q^0<0)$ and future infinity $(Q^0>0)$ in embedding space.}
    \label{fig::dsamb}
\end{figure}

$\left(d+1\right)$-dimensional de Sitter space with curvature radius $L$ can be realised as a co-dimension one hypersurface embedded in $\mathbb{R}^{1,d+1}$:
\begin{equation}\label{dShyp}
    \eta_{AB}X^{A}X^{B} = L^2,
\end{equation}
with
\begin{equation}
    \eta_{AB} = (-1,\underbrace{1,\ldots,1}_{d+1}).
\end{equation}
See figure \ref{fig::dsamb}. By construction this is a maximally symmetric space with isometry group $O\left(1,d+1\right)$, which has four disconnected components. The component that is continuously connected to the identity is $SO^{+}\left(1,d+1\right)$, whose elements preserve both space orientation and direction of time. Discrete elements include time reversal:
\begin{equation}
    T = \text{diag}(-1,\underbrace{1,\ldots,1}_{d+1}),
\end{equation}
and spatial reflection 
\begin{equation}\label{p}
    S = \text{diag}(1,-1,\underbrace{1,\ldots,1}_{d}).
\end{equation}
We will also be interested in the element
\begin{equation}\label{ap}
    A = \text{diag}(\underbrace{-1,\ldots,-1}_{d+2}),
\end{equation}
which is the antipodal transformation, which sends a point $x \in\,$dS$_{d+1}$ to the antipodal point ${\bar x}$: $X({\bar x}) = -X\left(x\right) $. This is conjugate to time reversal $T$ under $SO^{+}\left(1,d+1\right)$.

\vskip 4pt
Global coordinates corresponds to parameterising $X^A$ as
\begin{equation}
    X^A=L(\sinh \tau, \cosh \tau\,n_1\,\ldots,\cosh \tau\,n_{(d+1)}), \qquad \sum^{d+1}\limits_{i=1}n^2_i=1.
\end{equation}
The metric in these coordinates reads
\begin{equation}
    ds^2=L^2(-d\tau^2+\cosh^2 \tau \,d\Omega^2_d),
\end{equation}
where $\tau \in \left(-\infty,\infty\right)$ and $d\Omega^2_d$ is the metric for the $d$-dimensional unit sphere. 

\vskip 4pt
Another solution to \eqref{dShyp} is based on the parameterisation
\begin{subequations}
 \begin{align}
    X^0&=L\sinh\left(\frac{t_+}{L}\right)+\frac{\vec{x}^2_+}{2L}\, e^{\frac{t_+}{L}},\\
    X^i&= \frac{x^i_+}{L}\,e^{\frac{t_+}{L}},\quad i=1,\ldots,d, \\
    X^{d+1}&=-L\cosh\left(\frac{t_+}{L}\right)+\frac{\vec{x}^2_+}{2L}\, e^{\frac{t_+}{L}}.
\end{align}   
\end{subequations}
The metric in this case reads
\begin{equation}
    ds^2=-dt^2_++e^{\frac{2t_+}{L}}d\vec{x}^2_+.
\end{equation}
This coordinate system is referred to as the Expanding Poincar\'e Patch (EPP), and only covers half of the dS space-time, since in this parameterisation $X^0\geq X^{d+1}$. The other half with $X^0\leq X^{d+1}$ is referred to as the Contracting Poincar\'e Patch (CPP), which corresponds to the parameterisation
\begin{subequations}
 \begin{align}
    X^0&=L\sinh\left(\frac{t_-}{L}\right)-\frac{\vec{x}^2_-}{2L}\, e^{-\frac{t_-}{L}},\\
    X^i&= \frac{x^i_-}{L}\,e^{-\frac{t_-}{L}},\quad i=1,\ldots,d, \\
    X^{d+1}&=L\cosh\left(\frac{t_-}{L}\right)-\frac{\vec{x}^2_-}{2L}\, e^{-\frac{t_-}{L}},
\end{align}   
\end{subequations}
with metric
\begin{equation}
    ds^2=-dt^2_-+e^{-\frac{2t_-}{L}}d\vec{x}^2_-.
\end{equation}
In both the expanding and contracting patches it is useful to work in conformal time, in terms of which both patches have the same metric:
\begin{equation}\label{ppmetric}
    ds^2=\frac{L^2}{\eta^2_{\pm}}(-d\eta^2_\pm+d\vec{x}^2_\pm), \qquad \eta_\pm = L e^{\mp \frac{t_\pm}{L}}.
\end{equation}
\begin{figure}[htb]
    \centering
    \includegraphics[width=0.55\textwidth]{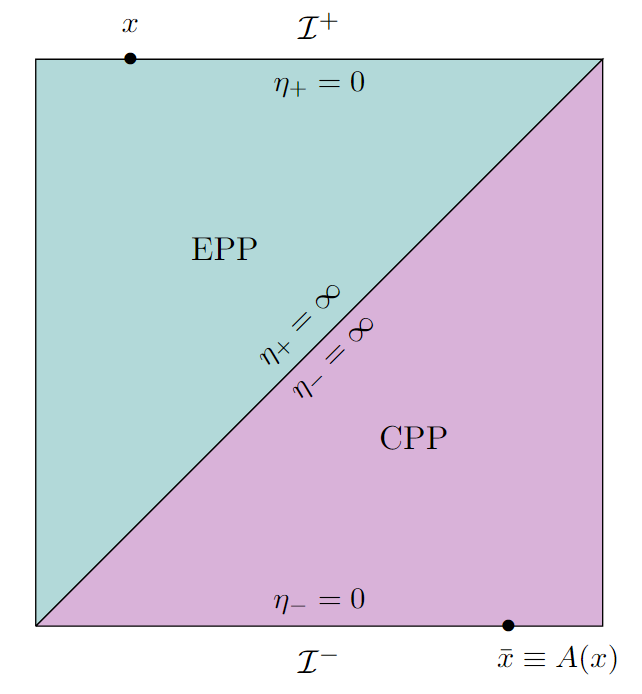}
    \caption{The Penrose diagram for de Sitter space, showing the Expanding and Contracting Poincar\'e patches and the locations of the conformal boundaries at past and future infinity. The antipodal map takes a point on the boundary at future infinity $\mathcal{I}^+$ to its opposite point on the boundary at past infinity $\mathcal{I}^-$.}
    \label{fig::dspen}
\end{figure}
The difference is that in the EPP the conformal time $\eta_+$ runs from $\eta_+=\infty$ at past infinity ($t_+=-\infty$) to $\eta_+=0$ at future infinity ($t_+=+\infty$) and in the CPP the conformal time $\eta_-$ runs from $\eta_-=0$ at past infinity ($t_-=-\infty$) to $\eta_-=\infty$ at future infinity ($t_-=+\infty$). Both the EPP and CPP can be covered simultaneously with metric \eqref{ppmetric} by making the replacements $\vec{x}_\pm \to \vec{x}$ and $\eta_\pm \to \eta \in (-\infty,+\infty)$, where the EPP is covered by $\eta = -\eta_+ <0$ and $\eta = \eta_- >0$ the CPP. In embedding coordinates this reads:
\begin{subequations}\label{dspoinc}
 \begin{align}
    X^0&=\frac{L}{2(-\eta)}\left(1-\eta^2+\vec{x}^2\right),\\
    X^i&=\frac{L}{2(-\eta)}x^i,\\
    X^{d+1}&=\frac{L}{2(-\eta)}\left(1+\eta^2-\vec{x}^2\right).
\end{align}   
\end{subequations}
In this parameterisation the antipodal transformation acts simply as:
\begin{equation}
A:x\longmapsto{\bar x} = (-\eta,\vec{x}), \quad \text{where} \quad x = (\eta, \vec{x}).
\end{equation}
\begin{figure}[htb]
    \centering
    \includegraphics[width=0.45\textwidth]{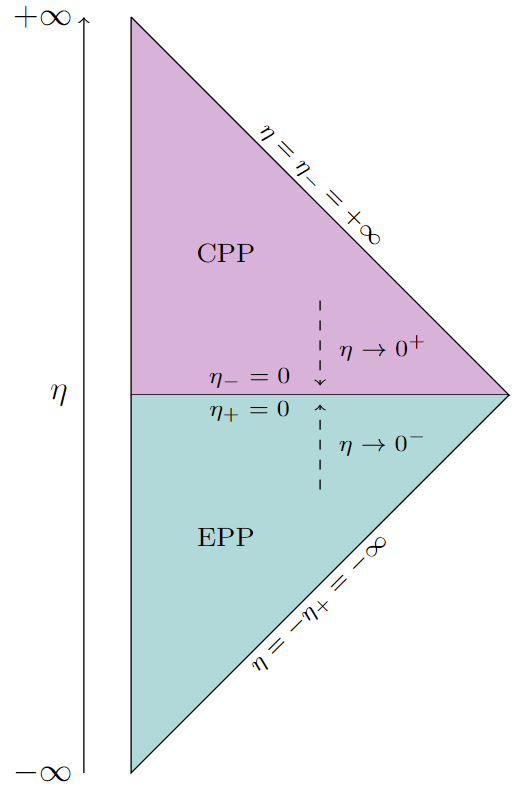}
    \caption{The EPP and CPP can simultaneously be covered by the time coordinate $\eta\in(-\infty,+\infty)$. The boundary at future infinity is reached from the EPP by sending $\eta\rightarrow0^-$, and the boundary at past infinity is reached from the CPP by sending $\eta\rightarrow0^+$.}
    \label{fig::dspen}
\end{figure}

\vskip 4pt
The conformal boundaries (past and future infinity) of dS space are identified with the projective null cone:
\begin{equation}
Q^2=0, \qquad Q \equiv \lambda Q, \qquad \lambda \in \mathbb{R}^+,
\end{equation}
where for the past and future boundary we have $Q^0 <0$ and $Q^0 >0$, respectively. In Poincar\'e coordinates the future boundary is reached by sending $\eta \to 0^-$ (i.e. in the EPP) with parameterisation
\begin{equation}\label{qp}
Q_+ = \left(\frac{1+\vec{x}^2}{2},\,\vec{x}\,,\frac{1-\vec{x}^2}{2}\right).
\end{equation}
Similarly, the past boundary is reached by sending $\eta \to 0^+$ through the CPP with parameterisation
\begin{equation}\label{qm}
Q_- = -\left(\frac{1+\vec{x}^2}{2},\,\vec{x}\,,\frac{1-\vec{x}^2}{2}\right),
\end{equation}
$Q_+$ on the future boundary is therefore related to $Q_-$ on the past boundary via antipodal transformation \eqref{ap}.

\vskip 4pt
The subgroup $SO^+\left(1,d+1\right)$ of the de Sitter isometry group acts on the boundaries at infinity like the connected component of the conformal group \cite{Costa:2011mg,Penedones:2016voo}. The global conformal group is isomorphic to $O\left(1,d+1\right)/\left\{\pm 1\right\}$ (see e.g. section 2.2 of \cite{Schottenloher:2008zz}) and discrete elements include the inversion transformation, which in the embedding formalism corresponds to the transformation $Q^{d+1} \to -Q^{d+1}$ or equivalently $Q^+ \leftrightarrow Q^-$ (see e.g. section 2.3.3 of \cite{Rychkov:2016iqz}). This is conjugate under  $SO^+\left(1,d+1\right)$ to the spatial reflection \eqref{p}. Note that the antipodal transformation \eqref{ap} instead is equivalent to the identity from the perspective of the global conformal group.

\section{Scalar Fields in the Expanding Poincar\'e Patch}
\label{sec::SFEPP}

In this section we review the relevant aspects of quantum scalar field theory in the expanding Poincar\'e patch of de Sitter space. For more complete and more detailed pedagogical reviews see e.g. \cite{Spradlin:2001pw,Baumann:2009ds,Anninos:2012qw,Akhmedov:2013vka}.

\vskip 4pt
Consider a free scalar field $\phi$ of generic mass with equation of motion
\begin{equation}\label{KG}
\left(\nabla^2-m^2\right)\phi = 0,
\end{equation}
where $\nabla^2$ is the Laplacian on dS$_{d+1}$.  The behaviour of the field as it approaches the boundary is especially simple, and in Poincar\'e coordinates reads:
\begin{equation}
\phi\left(\eta \to 0^-, \vec{x}\right)={\cal O}_+\left(\vec{x}\right)(-\eta)^{\Delta_+}+{\cal O}_-\left(\vec{x}\right)(-\eta)^{\Delta_-},
\end{equation} 
where $\Delta_++\Delta_-=d$, and under the connected component $SO^+(1,d+1)$ of the dS isometry group the ${\cal O}_\pm(\vec{x})$ transform as scalar primary operators with scaling dimension $\Delta_\pm$ in a $d$-dimensional Euclidean Conformal Field Theory. The scaling dimensions are fixed in terms of the mass as 
\begin{align}
m^2&=\Delta_+ \Delta_-,
\end{align}
which is convenient to parameterise via $\Delta_\pm=\frac{d}{2}\pm i
\nu$. The non-tachyonic irreducible representations of the de Sitter isometry group are \cite{Joung:2006gj,Joung:2007je}:
\begin{itemize}
    \item Principal Series: $\nu \in \mathbb{R}$, \: $m^2 \geq \left(\tfrac{d}{2}\right)^2$, \: \text{Massive Particles}.
    \item Complementary Series: $\nu \to -i\mu$, \: $\mu \in \left(0,\tfrac{d}{2}\right)$, \: $0 <m^2 < \left(\tfrac{d}{2}\right)^2$, \: \text{Light Particles}.
\end{itemize}
Massless particles correspond to $\mu = \tfrac{d}{2}$ and lie on the boundary of the complementary series.

\vskip 4pt
In Fourier space the mode expansion of $\phi$ takes the form\footnote{We normalise the mode functions with respect to the Klein-Gordon inner product: Defining $\phi_1\left(\eta,\vec{x}\right) =\int \frac{{\rm d}^dk}{\left(2\pi\right)^d}e^{+i\vec{k}\cdot \vec{x}}{\bar f}_{\vec{k}}\left(\eta\right)$ and $\phi_2\left(\eta,\vec{x}\right) =\int \frac{{\rm d}^dk}{\left(2\pi\right)^d}e^{+i\vec{k}\cdot \vec{x}}f_{\vec{k}}\left(\eta\right)$, we require
\begin{equation}
\left(\phi_1,\phi_1\right)=1, \quad \left(\phi_2,\phi_2\right)=-1, \quad \left(\phi_1,\phi_2\right)=0,
\end{equation}
where the Klein-Gordon inner product of functions $f$ and $g$ on a spatial slice $\Sigma$ with induced metric $\gamma_{ij}$ is defined via
\begin{equation}\label{KGINP}
\left(f,g\right)=-i \int_{\Sigma}{\rm d}^dx\,\sqrt{\gamma} n^\mu \left(f \partial_\mu g^* - g^* \partial_\mu f\right),
\end{equation}
where $n^\mu$ is a time-like unit vector normal to $\Sigma$.
}
\begin{equation}\label{modeexp}
\phi_{\vec{k}}\left(\eta\right) = f_{\vec{k}}\left(\eta\right) a^\dagger_{\vec{k}}\,+{\bar f}_{\vec{k}}\left(\eta\right)a_{-\vec{k}}\,, 
\end{equation}
where
\begin{equation}
    \phi\left(\eta,\vec{x}\right) = \int \frac{{\rm d}^dk}{\left(2\pi\right)^d}e^{+i\vec{k}\cdot \vec{x}}\phi_{\vec{k}}\left(\eta\right),
\end{equation}
and $a_{\vec{k}}$ and $a^\dagger_{\vec{k}}$ obey the usual commutation relations
\begin{equation}
[a_{\vec{k}_1},a^\dagger_{\vec{k}_2}] = \left(2\pi\right)^d \delta(\vec{k}_1-\vec{k}_2), \qquad [a_{\vec{k}_1},a_{\vec{k}_2}] =[a^\dagger_{\vec{k}_1},a^\dagger_{\vec{k}_2}]=0\,.
\end{equation}

\vskip 4pt
The vacuum state $| \Omega \rangle$ can be defined in the usual way via
\begin{equation}
a_{\vec{k}} | \Omega \rangle=0, \quad \forall\, \vec{k},
\end{equation}
i.e. that it is annihilated by all annihilation operators. The vacuum state thus depends on the choice of modes in the expansion \eqref{modeexp}. The Wightman function of a free massive scalar can be used to characterise the various de Sitter invariant vacua. Consider the Wightman two-point function
\begin{equation}
G_{\Omega}\left(x_1;x_2\right)=\langle \Omega | \phi\left(x_1\right)\phi\left(x_2\right) | \Omega \rangle,
\end{equation} 
which characterises the vacuum state $| \Omega \rangle$. This can only depend on the two-point invariant
\begin{equation}
s\left(x_1,x_2\right)=X_1\left(x_1\right) \cdot X_2\left(x_2\right),
\end{equation}
which is convenient to express through the variable
 \begin{align}
\sigma\left(x_1,x_2\right)&=\frac{1+X_1\left(x_1\right)\cdot X_2\left(x_2\right)}{2},
\end{align}   
where $\sigma\left(x,x\right)$=1, $\sigma\left(x,{\bar x}\right)=0$ and in the second equality we gave its form in Poincar\'e coordinates \eqref{dspoinc}.
In terms of $\sigma$ the Klein-Gordon equation \eqref{KG} for the Wightman function reads \cite{Allen:1985ux}
\begin{equation}\label{propeq}
[\sigma\left(1-\sigma\right)\partial^2_{\sigma}G_{\Omega}\left(\sigma\right)-\left(\tfrac{d+1}{2}\right)\left(2\sigma-1\right)\partial_{\sigma}G_{\Omega}\left(\sigma\right)]-m^2G_{\Omega}\left(\sigma\right)=0,
\end{equation}
which is Euler's hypergeometric differential equation. This has two independent solutions,
\begin{equation}\label{gensolw}
G_{\Omega}\left(\sigma\right) = A\,{}_2F_1(\tfrac{d}{2}+i\nu,\tfrac{d}{2}-i\nu;\tfrac{d+1}{2};\sigma)+B\,{}_2F_1(\tfrac{d}{2}+i\nu,\tfrac{d}{2}-i\nu;\tfrac{d+1}{2};1-\sigma),
\end{equation}
which are related under the antipodal transformation
\begin{equation}
\sigma(x,{\bar y}) = 1-\sigma(x,y),
\end{equation}
which is a symmetry of the differential equation \eqref{propeq}. The solution with $B=0$ has a singularity when $\sigma(x,y)=1$, which is the usual singularity that occurs when $x$ and $y$ are null separated. The solution with $A=0$ instead has a singularity when $\sigma(x,y)=0$, which occurs when $x$ and $y$ are antipodal. The latter singularity is unusual since antipodal points in dS are separated by a horizon.

\vskip 4pt
There is therefore an infinite one-parameter family of de Sitter invariant Wightman functions corresponding to linear combinations of solutions \eqref{gensolw}. To each Wightman function in this family there is a corresponding de Sitter invariant vacuum state which, as we will review in the following, are parameterised by a superselection parameter $\alpha$ \cite{Chernikov:1968zm,Burges:1984qm,Mottola:1984ar,Allen:1985ux}.

\subsection{Bunch-Davies vacuum}

A generic de Sitter invariant vacuum differs from the usual Minkowski vacuum owing to the fact that, in addition to the usual short distance singularity, there is also an antipodal singularity in the solution \eqref{gensolw} for $B\ne 0$. The solution with $B=0$ therefore identifies a unique vacuum with the same light-cone singularities as in Minkowski space (the Hadamard condition), which is known as the Bunch-Davies vacuum \cite{Chernikov:1968zm,Schomblond:1976xc,Gibbons:1977mu,Bunch:1978yq}. The coefficient $A$ in the Wightman function and the $i\epsilon$ prescription for the branch cut at $\sigma \in \left(1,\infty\right)$ are fixed by requiring the same short distance singularity as in flat space, giving:
\begin{subequations}\label{bdw2pt}
 \begin{align}
G_{W}^{(0)}\left(x_1;x_2\right) &=G^{(0)}\left(\sigma-i \epsilon\, \text{sgn}(\eta_1-\eta_2)\right),\\
G^{(0)}\left(\sigma\right)&= \frac{\Gamma\left(\tfrac{d}{2}+i\nu\right)\Gamma\left(\tfrac{d}{2}-i\nu\right)}{\left(4\pi\right)^{\frac{d+1}{2}}\Gamma\left(\frac{d+1}{2}\right)}\,{}_2F_1(\tfrac{d}{2}+i\nu,\tfrac{d}{2}-i\nu;\tfrac{d+1}{2};\sigma),
\end{align}   
\end{subequations}
where we introduced the superscript ``$(0)$" to denote the Bunch-Davies vacuum.

\vskip 4pt
The $i \epsilon$ prescriptions for the time-ordered $(T)$ and anti-time-ordered $({\bar T})$ propagators are instead:
\begin{subequations}\label{Gtbat}
\begin{align}
G^{(0)}_{T}\left(\sigma\right) &=G^{(0)}\left(\sigma-i \epsilon\right),\\
G_{{\bar T}}^{(0)}\left(\sigma\right) &=G^{(0)}\left(\sigma+i \epsilon\right).
\end{align}     
\end{subequations}
In Poincar\'e coordinates these can be equivalently written as:
\begin{subequations}
 \begin{align}
G^{(0)}_{T}\left(\eta_1;\eta_2\right) &=\theta(\eta_1-\eta_2)G^{(0)}_{W}\left(\eta_1;\eta_2\right)+\theta(\eta_2-\eta_1)G^{(0)}_{W}\left(\eta_2;\eta_1\right),\\
G_{{\bar T}}^{(0)}\left(\eta_1;\eta_2\right) &=\theta(\eta_1-\eta_2)G^{(0)}_{W}\left(\eta_2;\eta_1\right)+\theta(\eta_2-\eta_1)G^{(0)}_{W}\left(\eta_1;\eta_2\right),
\end{align}    
\end{subequations}
in terms of the Heaviside step function $\theta\left(x\right)$.

\vskip 4pt
It is often useful to decompose dS two-point functions in the Bunch-Davies vacuum according to
\begin{equation}
    G^{(0)}\left(\sigma\right) =  G^{(0)}_{\Delta_+}\left(\sigma\right)+G^{(0)}_{\Delta_-}\left(\sigma\right),
\end{equation}
where 
\begin{equation}
    G^{(0)}_{\Delta}\left(\sigma\right) = C^{\text{dS}}_{\Delta}\left(-4 \sigma\right)^{-\Delta}{}_2F_1\left(\begin{matrix}\Delta,\Delta-\frac{d}{2}+\frac{1}{2}\\2\Delta-d+1\end{matrix},\frac{1}{\sigma}\right).
\end{equation}
In the case that one of the two points lies on the future or past boundary one can then expand \eqref{Gtbat} for large $s$, giving:
\begin{align}
G^{(0)}_{T\,({\bar T})}\left(s \to \infty \right)= K^{(0)}_{\Delta_+,\, T\,({\bar T})}\left(s\right) + K^{(0)}_{\Delta_-,\, T\,({\bar T})}\left(s\right),
\end{align}
which naturally identifies time-ordered and anti-time-ordered bulk-to-boundary propagators:
\begin{subequations}\label{bdbubo}
 \begin{align}
K^{(0)}_{\Delta,\, T}\left(s\right) = \frac{C^{\text{dS}}_{\Delta}}{(-2s+i\epsilon)^{\Delta}},\\
K^{(0)}_{\Delta,\, {\bar T}}\left(s\right) = \frac{C^{\text{dS}}_{\Delta}}{(-2s-i\epsilon)^{\Delta}},
\end{align}   
\end{subequations}
with definite scaling dimension $\Delta$ and two-point coefficient: 
\begin{equation}\label{dS2ptcoeff}
C^{\text{dS}}_{\Delta} = \frac{1}{4 \pi^{\frac{d+2}{2}}} \Gamma\left(\Delta\right)\Gamma(\tfrac{d}{2}-\Delta).
\end{equation}

\vskip 4pt
In the Bunch-Davies vacuum the mode functions $f^{(0)}_{\vec{k}}$ and ${\bar f}^{(0)}_{\vec{k}}$ are given by Hankel functions of the second and first kind respectively:
\begin{subequations}\label{bdmodf}
 \begin{align}
f^{(0)}_{\vec{k}}\left(\eta\right)&=\left(-\eta\right)^{\frac{d}{2}} \frac{\sqrt{\pi}}{2} e^{\frac{\pi \nu}{2}}H^{(2)}_{i\nu}\left(-k \eta\right),\\
{\bar f}^{(0)}_{\vec{k}}\left(\eta\right)&=\left(-\eta\right)^{\frac{d}{2}} \frac{\sqrt{\pi}}{2} e^{-\frac{\pi \nu}{2}}H^{(1)}_{i\nu}\left(- k \eta\right),
\end{align}   
\end{subequations}
where the overall normalisation is fixed by the Klein-Gordon inner product \eqref{KGINP}. In the Bunch-Davies vacuum these are simply related under antipodal transformation:
\begin{equation}\label{bdap}
    {\bar f}^{(0)}_{\vec{k}}\left(\eta\right) = -\,e^{\frac{i\pi d}{2}}f^{(0)}_{\vec{k}}(\eta\, e^{-i \pi}).
\end{equation}
In terms of the mode functions the Wightman function is given simply by: 
\begin{equation}
    G^{(0)}_W(\eta_1,\vec{k}_1;\eta_2,\vec{k}_2)={\bar f}_{\vec{k}_1}\left(\eta_1\right)f_{\vec{k}_2}\left(\eta_2\right).
\end{equation}

\subsection{Bogoliubov states and $\alpha$-vacua}
\label{subsec::alphavacua}

The mode functions for a general Bogoliubov state are a linear combination of the Bunch-Davies mode functions \eqref{bdmodf}: 
\begin{equation}\label{amodes}
f^{(\alpha)}_{\vec{k}}(\eta) = A_k\, f^{(0)}_{\vec{k}}(\eta)+B_k\, {\bar f}^{(0)}_{\vec{k}}(\eta),
\end{equation}
where orthonormality \eqref{KGINP} of the mode functions requires
\begin{equation}
|A_k|^2-|B_k|^2=1.
\end{equation}
This equation is solved by 
\begin{equation}
A_k = \cosh \alpha_k, \qquad B = e^{i\beta_k}  \sinh \alpha_k,
\end{equation}
with $\alpha_k \in \left[0,\infty\right)$ and $\beta_k \in \left[0,2\pi\right)$, where we have neglected a possible overall phase that does not contribute to expectation values. The Wightman function is invariant under the connected component of the dS isometry group only if $\alpha_k, \beta_k$ are independent of $k$, and the full disconnected isometry group $O(1,d+1)$ only if furthermore $\beta = 0$ \cite{Allen:1985ux}.

\vskip 4pt
The annihilation operator associated to the mode basis \eqref{amodes} is then given in terms of the Bunch-Davies creation and annihilation operators $a^{(0)}_{\vec{k}}{}^\dagger$ and $a^{(0)}_{\vec{k}}$ via
\begin{equation}
    a^{(\alpha)}_{\vec{k}} = \cosh \alpha_k\, a^{(0)}_{\vec{k}}-e^{i\beta_k}  \sinh \alpha_k\, a^{(0)}_{-\vec{k}}{}^\dagger.
\end{equation}
This defines a vacuum state via
\begin{equation}\label{avac}
    a^{(\alpha)}_{\vec{k}} \ket{\alpha} = 0,
\end{equation}
with the Bunch-Davies vacuum (i.e. $\alpha_k=\beta_k=0$) then denoted by $\ket{0}$. For momentum-independent $\alpha_k$ and $\beta_k$ (i.e. $\alpha_k,\,\beta_k \to \alpha, \beta$), the states $\ket{\alpha}$ and $\ket{0}$ do not lie in the same Hilbert space and $\alpha$ is a superselection parameter. In other words, in this case each $\alpha$ and $\beta$ define the de Sitter invariant ground state of a different Hilbert space. If $\alpha_k, \beta_k$ are instead momentum dependent, this dependence can be chosen such that $\ket{\alpha}$ and $\ket{0}$ lie in the same Hilbert space -- in which case the $\ket{\alpha}$ are excited (squeezed) states in the Fock space constructed over the Bunch-Davies vacuum \cite{Einhorn:2003xb,deBoer:2004nd}.

\vskip 4pt
The Wightman function in a generic Bogoliubov state \eqref{avac} in momentum space can therefore be expressed in terms of the Bunch-Davies Wightman function \eqref{bdw2pt} as:
\begin{multline}\label{wafrombd}
G^{(\alpha)}_{W}(\eta_1;\eta_2) = \cosh^2\alpha_k \, G^{(0)}_{W}(\eta_1;\eta_2)+\sinh^2\alpha_k \, G^{(0)}_{W}({\bar \eta}^{-}_1;{\bar \eta}^+_2)\\-\frac{1}{2}\sinh 2\alpha_k [ e^{i\left(\beta_k+\frac{\pi d}{2}\right)} \, G^{(0)}_{W}(\eta_1;{\bar \eta}^+_2)+ e^{-i\left(\beta_k+\frac{\pi d}{2}\right)} \, G^{(0)}_{W}({\bar \eta}^-_1;\eta_2)], 
\end{multline}
where we used the property \eqref{bdap} of Bunch-Davies mode functions under antipodal transformation and we introduced the notation ${\bar \eta}^{\pm}=e^{\mp i\pi}\eta$. For $\alpha$-vacua, using that $\sigma({\bar x},{\bar y})=\sigma(x,y)$, $\sigma(x,y)=\sigma(y,x)$ and $\sigma({\bar x},y)=1-\sigma(x,y)$, the above expression for the Wightman function reduces to the form \eqref{gensolw} in position space, allowing us to relate the coefficients $A,\, B$ to $\alpha,\, \beta$.

\vskip 4pt
In turn, the time-ordered and anti-time-ordered propagators can be expressed in terms of those in the Bunch-Davies vacuum (see appendix \ref{app::PI}): 
\begin{subequations}\label{TaTBD}
\begin{align}
G^{(\alpha)}_{T}(\eta_1;\eta_2)&=P^+_{\Delta} G^{(0)}_{\Delta_+,T}(\eta_1;\eta_2)+e^{2\Delta_+ \pi i}P^-_{\Delta} G^{(0)}_{\Delta_+,T}({\bar \eta}^+_1;{\bar \eta}^+_2) + (\Delta_+ \to \Delta_-), \\
G^{(\alpha)}_{{\bar T}}(\eta_1;\eta_2)&=M^-_{\Delta} G^{(0)}_{\Delta_+,{\bar T}}(\eta_1;\eta_2)+ \, e^{-2\Delta_+ \pi i} M^+_{\Delta} G^{(0)}_{\Delta_+,{\bar T}}({\bar \eta}^-_1;{\bar \eta}^-_2)+ (\Delta_+ \to \Delta_-), 
\end{align}    
\end{subequations}
where
\begin{subequations} 
 \begin{align} \nonumber
  \hspace*{-0.35cm}  P^+_{\Delta} &= \left(\cosh^2\alpha_k-\frac{1}{2} e^{-i \beta_k }\sinh 2\alpha_k e^{- \pi \nu} \right), \quad
P^-_{\Delta}=\left(\sinh^2\alpha_k-\frac{1}{2} e^{+ i \beta_k }\sinh 2\alpha_k e^{+ \pi \nu} \right),\\ \nonumber
\hspace*{-0.35cm} M^+_{\Delta} &=\left(\sinh^2\alpha_k-\frac{1}{2} e^{-i \beta_k }\sinh 2\alpha_k e^{-\pi \nu} \right), \quad
M^-_{\Delta} = \left(\cosh^2\alpha_k-\frac{1}{2} e^{+ i \beta_k }\sinh 2\alpha_k e^{+ \pi \nu} \right).
\end{align}   
\end{subequations}

Sending one of the bulk points to the future boundary one obtains analogous expressions for the bulk-to-boundary propagators in terms of their Bunch-Davies counterparts \eqref{bdbubo} (see appendix \ref{app::PI}): 
\begin{subequations}\label{buboafrombd}
\begin{align}
K^{(\alpha)}_{\Delta,\, T}\left(\eta,k\right) &= P^+_{\Delta} K^{(0)}_{\Delta,\, T}(\eta,k) + P^-_{\Delta} e^{\Delta \pi i}K^{(0)}_{\Delta,\, T}({\bar \eta}^+,k),\\
K^{(\alpha)}_{\Delta,\, {\bar T}}(\eta,k) &= M^+_{\Delta} e^{-\Delta \pi i}K^{(0)}_{\Delta,\, {\bar T}}({\bar \eta}^-,k) + M^-_{\Delta} K^{(0)}_{\Delta,\, {\bar T}}(\eta,k).
\end{align}    
\end{subequations}

\vskip 4pt
We see that two-point functions in a generic Bogoliubov state can be expressed in terms of corresponding two-point functions in the Bunch-Davies vacuum ($\alpha_k=\beta_k=0$). This hints that it might be possible to reformulate the Feynman rules for late-time correlators in a generic Bogoliubov initial state in terms of their counterparts in the Bunch-Davies vacuum and, in turn, in terms of those for boundary correlators in EAdS via analytic continuation as understood in \cite{Sleight:2019hfp,Sleight:2020obc,Sleight:2021plv}. This will be made precise in the following section.

\section{Late-time correlation functions}
\label{sec::LTCF}
In this work we are interested in correlation functions on the future boundary of de Sitter space, which can be calculated using the ``in-in formalism" \cite{Maldacena:2002vr,Bernardeau:2003nx,Weinberg:2005vy} (for a review see e.g. \cite{Chen:2017ryl}). The in-in formalism prescribes that the correlation function of some set of fields on a given time slice at time $\eta$ is given by
\begin{multline}\label{inincorr}
\langle \phi(\eta,\vec{x}_1)  \ldots  \phi(\eta,\vec{x}_n) \rangle \\
= \langle \Omega(\eta_0)|\, {\bar T}\left(\exp[i \int^\eta_{\eta_0}{\rm d}\eta^\prime H]\right) \phi(\eta,\vec{x}_1)  \ldots  \phi(\eta,\vec{x}_n) T\left(\exp[-i \int^\eta_{\eta_0}{\rm d}\eta^\prime H]\right)|\Omega(\eta_0)\rangle.
\end{multline}
One starts at initial time $\eta_0=-\infty$ with vacuum $|\Omega(\eta_0)\rangle$ in the interacting theory, evolves in a time ordered ($T$) fashion with respect to Hamiltonian $H$ to the time $\eta$ and then evolves back in an anti-time-ordered fashion (${\bar T}$).
To compute such correlators perturbatively it is useful to work in the interaction picture, where one splits the Hamiltonian into a free and interacting part $H = H_0 + H_{\text{int}}$ and the correlator \eqref{inincorr} can be re-written as:
\begin{multline}
\langle \phi(\eta,\vec{x}_1)  \ldots  \phi(\eta,\vec{x}_n) \rangle \\
= \frac{{}_0\langle 0|\, {\bar T}\left(\exp[i \int^\eta_{-\infty}{\rm d}\eta^\prime H_{\text{int}}(\phi_0)]\right) \phi_0(\eta,\vec{x}_1)  \ldots  \phi_0(\eta,\vec{x}_n) T\left(\exp[-i \int^\eta_{-\infty}{\rm d}\eta^\prime H_{\text{int}}(\phi_0)]\right)|0\rangle_0}{{}_0\langle 0|\, {\bar T}\left(\exp[i \int^\eta_{-\infty}{\rm d}\eta^\prime H_{\text{int}}(\phi_0)]\right)T\left(\exp[-i \int^\eta_{-\infty}{\rm d}\eta^\prime H_{\text{int}}(\phi_0)]\right)|0\rangle_0},
\end{multline}
 where $\ket{0}_0$ is the free theory vacuum and and the $\phi_0$ evolve according to the free theory Hamiltonian $H_0$. At late times we have $\eta \sim 0$. In this work we take $\ket{0}_0$ to be a Bogoliubov state \eqref{avac}. The perturbative expansion is generated by expanding the exponentials so that the correlator can then be computed perturbatively in the usual way using Wick contractions. This gives rise to four bulk-to-bulk propagators, since two fields in a given Wick contraction can be either time-ordered (+) or anti-time-ordered ($-$):
 \begin{subequations}\label{ininbubu}
  \begin{align}
 G_{++}(x_1;x_2)&=G_{T}(x_1;x_2),\\
 G_{+-}(x_1;x_2)&=G_{W}(x_2;x_1),\\
 G_{-+}(x_1;x_2)&=G_{W}(x_1;x_2),\\
 G_{--}(x_1;x_2)&=G_{{\bar T}}(x_1;x_2),
 \end{align}    
 \end{subequations}
 and two bulk-to-boundary propagators:
 \begin{subequations}\label{ininbubo}
 \begin{align}
 K_{\Delta,\, +}(x,\vec{y}) &= K_{\Delta,\, T}(x,\vec{y}),\\
 K_{\Delta,\, -}(x,\vec{y}) &= K_{\Delta,\, {\bar T}}(x,\vec{y}).
 \end{align}     
 \end{subequations}

\subsection{Late-time correlators from Bunch-Davies}
\label{subsec::alphafromBD}

We have seen that two-point functions in a generic Bogoliubov state can be expressed in terms of their Bunch-Davies counterparts. This in turn suggests that late-time correlation functions for Bogoliubov initial states might be obtained from the knowledge of their Bunch-Davies counterparts.\footnote{This has indeed been observed in some examples of three- and four-point diagrams involving massless scalars and gravitons \cite{Xue:2008mk,Jain:2022uja,Ghosh:2023agt,Ansari:2024pgq} in $\alpha$-vacua.} This is particularly clear in Fourier space, where the antipodal transformation of a bulk point in a Bunch-Davies two-point function can be replaced by a flip in the sign of the boundary momentum modulus (see appendix \ref{app::PI}):
\begin{subequations}\label{bubudSap}
 \begin{align}
    G^{(0)}_{\pm 
{\hat \pm}}({\bar \eta}^\pm_1,k_1;{\bar \eta}_2^{{\hat \pm}},k_2)&=(e^{\pm \pi i})^{\Delta_+-d}(e^{{\hat \pm} \pi i})^{\Delta_--d}G^{(0)}_{\pm 
{\hat \pm}}(\eta_1,{\bar k}_1^{\pm};\eta_2,{\bar k}_2^{{\hat \pm}}),\\
G^{(0)}_{\pm 
\mp}({\bar \eta}_1^{\pm},k_1;\eta_2,k_2)&=(e^{\pm \pi i})^{\Delta_+-d}G^{(0)}_{\pm 
{\hat \pm}}(\eta_1,{\bar k}_1^{\pm};\eta_2,k_2),\\
G^{(0)}_{\pm 
\mp}(\eta_1,k_1;{\bar \eta}_2^{\mp},k_2)&=(e^{\mp \pi i})^{\Delta_--d}G^{(0)}_{\pm 
\mp}(\eta_1,k_1;\eta_2,{\bar k}_2^{\mp}),
\end{align}   
\end{subequations}
and 
\begin{align}\label{bubodSap}
    K^{(0)}_{\Delta,\,\pm}({\bar \eta}^{\pm};k)&=(e^{\pm \pi i})^{\Delta-d}K^{(0)}_{\Delta,\,\pm}(\eta;{\bar k}^{\pm}).
\end{align}
The Schwinger-Keldysh propagators \eqref{ininbubu} and \eqref{ininbubo} for a generic Bogoliubov state can therefore be expressed in terms of their Bunch-Davies counterparts with rotated momenta via: 
\begin{subequations}\label{bubufromBD}
\begin{align}\nonumber
G^{(\alpha)}_{++}(\eta_1;\eta_2)&=P^+_{\Delta_+} G^{(0)}_{\Delta_+,++}(\eta_1,k;\eta_2,k)+e^{-2\nu\pi }P^-_{\Delta_+}  G^{(0)}_{\Delta_+,++}(\eta_1,{\bar k}^+;\eta_2,{\bar k}^+)\\& + (\Delta_+ \to \Delta_-),\\ \nonumber
G^{(\alpha)}_{--}(\eta_1;\eta_2)&=M^-_{\Delta_+} G^{(0)}_{\Delta_+,--}(\eta_1,k;\eta_2,k)+ \, e^{2\nu \pi }  M^+_{\Delta_+}G^{(0)}_{\Delta_+,--}(\eta_1,{\bar k}^-;\eta_2,{\bar k}^-)\\ &+ (\Delta_+ \to \Delta_-), \\ \nonumber
G^{(\alpha)}_{-+}(\eta_1;\eta_2)& = \cosh^2\alpha_k \, G^{(0)}_{\Delta_+,-+}(\eta_1,k;\eta_2,k)+\sinh^2\alpha_k \,e^{2\nu \pi}G^{(0)}_{\Delta_+,-+}(\eta_1,{\bar k}^-;\eta_2,{\bar k}^+)\\ \nonumber &-\frac{1}{2}\sinh 2\alpha_k\,e^{\nu \pi } [ e^{i\beta_k}\, G^{(0)}_{\Delta_+,-+}(\eta_1,k;\eta_2,{\bar k}^+)+ e^{-i\beta_k} \, G^{(0)}_{\Delta_+,-+}(\eta_1,{\bar k}^-;\eta_2,k)]\\ &+ (\Delta_+ \to \Delta_-), \\ \nonumber
G^{(\alpha)}_{+-}(\eta_1;\eta_2) &= \cosh^2\alpha_k \, G^{(0)}_{\Delta_+,+-}(\eta_1,k;\eta_2,k)+\sinh^2\alpha_k \,e^{-2\nu \pi } G^{(0)}_{\Delta_+,+-}(\eta_1,{\bar k}^+;\eta_2,{\bar k}^-)\\ \nonumber
 &\hspace*{-0.25cm}-\frac{1}{2}\sinh 2\alpha_k\,e^{-\nu\pi}[ e^{i\beta_k}\, G^{(0)}_{\Delta_+,+-}(\eta_1,k;\eta_2,{\bar k}^-)+ e^{-i\beta_k} \, G^{(0)}_{\Delta_+,+-}(\eta_1,{\bar k}^+;\eta_2,k)]\\ &+ (\Delta_+ \to \Delta_-), 
\end{align}    
\end{subequations}
and
\begin{subequations}
 \label{bubofromBD}
\begin{align}
K^{(\alpha)}_{\Delta,\, +}\left(\eta,k\right) &= P^+_{\Delta} K^{(0)}_{\Delta,\, +}\left(\eta,k\right) + P^-_{\Delta} e^{(2\Delta-d) \pi i}K^{(0)}_{\Delta,\, +}(\eta,{\bar k}^+),\\
K^{(\alpha)}_{\Delta,\, -}\left(\eta,k\right) &= M^+_{\Delta} e^{-(2\Delta-d) \pi i}K^{(0)}_{\Delta,\, -}(\eta,{\bar k}^-) + M^-_{\Delta} K^{(0)}_{\Delta,\, -}\left(\eta,k\right).
\end{align}   
\end{subequations}
These identities provide a proof, to all orders in perturbation theory, that late-time correlators involving scalar fields in a generic Bogoliubov initial state can be expressed in terms of their Bunch-Davies counterparts with the magnitudes of some of the boundary momenta flipped by $e^{\pm \pi i}$.

\vskip 4pt
In a series \cite{Sleight:2019hfp,Sleight:2020obc,Sleight:2021plv} of works it has been shown that the Schwinger-Keldysh formalism for late-time correlators in the Bunch-Davies vacuum of de Sitter space could be recast in terms of Feynman rules for boundary correlators in Euclidean anti-de Sitter space. The observation of this section, i.e. that the Schwinger-Keldysh formalism for late-time correlators for a Bogoliubov initial state can be reformulated in terms of Schwinger-Keldysh propagators in the Bunch-Davies vacuum, suggests that the connection to boundary correlators in EAdS might be extended to Bogoliubov initial states as well. This will be made concrete in the following section.

\subsection{Rotation to Euclidean AdS}
\label{subsec::toEAdS}

In this section we show that the Feynman rules for late-time correlators for a Bogoliubov initial state can be recast in terms of propagators for corresponding boundary correlators in Euclidean AdS. 

\begin{figure}[t]
    \centering
    \includegraphics[width=0.6\textwidth]{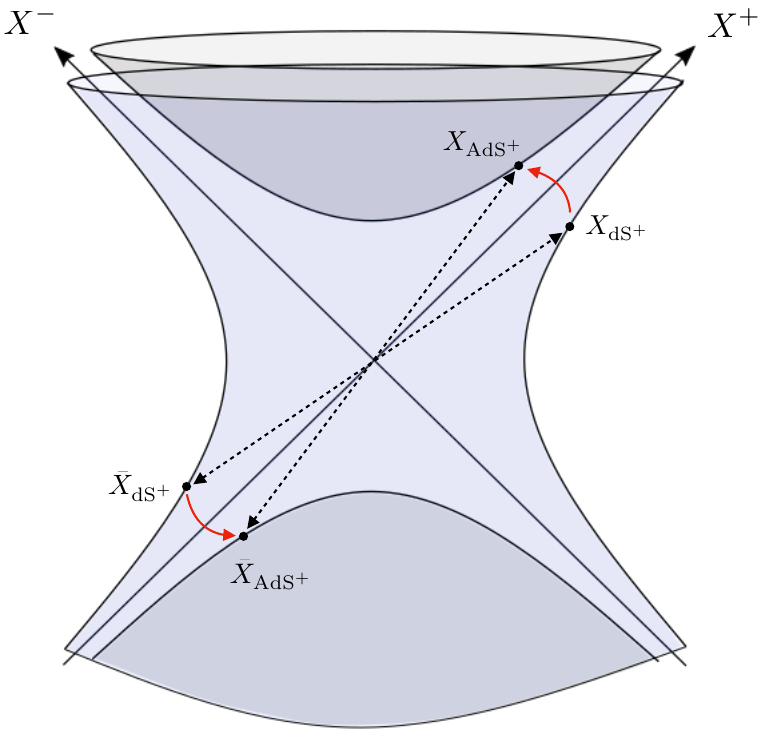}
    \caption{Points in the EPP analytically continue to the upper sheet of the EAdS hyperboloid while their antipodal points in the CPP analytically continue to the lower sheet.}
    \label{fig::ds2ads}
\end{figure}

\vskip 4pt
Consider $(d+1)$-dimensional Euclidean anti-de Sitter space (EAdS$_{d+1}$) with curvature radius $L=1$. This can be realised as the following co-dimension one hypersurface in $\mathbb{R}^{1,d+1}$:
\begin{equation}\label{EAdShyp}
    \eta_{AB}X^{A}X^{B} = -1,
\end{equation} 
and therefore, like dS$_{d+1}$, has isometry group $O(1,d+1)$. This is a two-sheeted hyperboloid (see figure \ref{fig::ds2ads}), which can be parameterised in Poincar\'e coordinates as 
\begin{align}
X_{\text{AdS}^\pm} = \pm \frac{1}{z}\left(\frac{1+z^2+\vec{x}^2}{2},\,\vec{x},\frac{1-z^2-\vec{x}^2}{2}\right), \qquad z>0,
\end{align}
where the upper sheet is parameterised by $X_{\text{AdS}^+}$ with $X_{\text{AdS}^+}^0 >0$ and the lower sheet by $X_{\text{AdS}^-}$ with $X^0_{\text{AdS}^-}<0$. For both sheets the boundary is reached upon sending $z \to 0$ and parameterised by $Q_\pm$ as in the Poincar\'e patches of dS space-time, equations \eqref{qp} and \eqref{qm}.

\vskip 4pt
Two-point functions are now functions of the EAdS chordal distance,
\begin{equation}
\sigma_{\text{AdS}}\left(x,y\right)=\frac{1+X_{\text{AdS}}\left(x\right)\cdot Y_{\text{AdS}}\left(y\right)}{2}.
\end{equation}
For a scalar field of mass 
\begin{equation}
m^2_{\text{AdS}} = -\Delta_+ \Delta_-,
\end{equation}
the bulk-to-bulk propagators are 
\begin{equation}\label{bubuads}
    G^{\text{AdS}}_{\Delta_\pm}\left(\sigma_{\text{AdS}}\right) = C^{\text{AdS}}_{\Delta_\pm} \left(-4\sigma_{\text{AdS}}\right)^{-\Delta_\pm} {}_2F_1\left(\Delta_\pm,\Delta_\pm-\frac{d}{2}+\frac{1}{2},2\Delta_\pm-d+1,\frac{1}{\sigma_{\text{AdS}}}\right),
\end{equation}
with corresponding bulk-to-boundary propagators:
\begin{align}\label{buboads}
   K^{\text{AdS}}_{\Delta}\left(s_{\text{AdS}}\right) &= \frac{C^{\text{AdS}}_{\Delta}}{\left(-2 s_{\text{AdS}} \right)^{\Delta}},
\end{align}
where $s_{\text{AdS}}(X,Q) = X\cdot Q$, and the normalisation is given by 
\begin{equation}\label{ads2pt}
C^{\text{AdS}}_{\Delta}=\frac{\Gamma\left(\Delta\right)}{2\pi^{\frac{d}{2}}\Gamma\left(\Delta-\frac{d}{2}+1\right)}.
\end{equation}
In Poincar\'e coordinates the bulk-to-boundary propagators read 
\begin{align}
    K^{\text{AdS}}_{\Delta}\left(z,\vec{x};\vec{y}\right)=C^{\text{AdS}}_{\Delta} \left(\frac{z}{z^2+\left(\vec{x}-\vec{y}\right)^2}\right)^{\Delta}.
\end{align}

\vskip 4pt
It is well known that dS and EAdS are formally related under analytic continuation $X_{\text{dS}} \to \pm i X_{\text{AdS}}$, which can be seen by comparing the embedding equations \eqref{dShyp} and \eqref{EAdShyp}. The upshot is that dS propagators are a linear combination of analytically continued propagators for a field of the same quantum numbers in EAdS. For the Bunch-Davies vacuum this was worked out in a series of works \cite{Sleight:2019mgd,Sleight:2020obc,Sleight:2021plv}, where it was shown that the Feynman rules for late-time correlators in the dS Bunch-Davies vacuum could be recast as Feynman rules for corresponding Witten diagrams in EAdS following the analytic continuations:
\begin{subequations}\label{poinanal}
 \begin{align}
    \emph{time-ordered points}: \quad &\eta\, \to\, +iz,\\
    \emph{anti-time-ordered points}: \quad &\eta\, \to\, -iz,
\end{align}   
\end{subequations}
which hold for both the expanding and contracting patches. The expanding patch continues to the upper sheet $X_{\text{AdS}^+}$ of EAdS and the contracting patch to the lower sheet $X_{\text{AdS}^-}$. In terms of embedding coordinates this reads:
\begin{subequations}
 \begin{align}
time\text{-}ordered\:points\,&: \quad X_{\text{dS}^\pm} \, \to \, \pm i X_{\text{AdS}^\pm},\\
anti\text{-}time\text{-}ordered\:points\,&: \quad X_{\text{dS}^\pm} \, \to \, \mp i X_{\text{AdS}^\pm}.
\end{align}   
\end{subequations}
For the Schwinger-Keldysh propagators, under the above analytic continuations we have \cite{Sleight:2020obc,Sleight:2021plv}:
\begin{subequations}\label{BDtoEAdS}
  \begin{align}
 G^{(0)}_{\pm {\hat \pm}}(X_{\text{dS}^+},Y_{\text{dS}^+})&\to c^{\text{dS-AdS}}_{\Delta_+}e^{\mp \frac{\Delta_+\pi i}{2}}e^{{\hat \mp} \frac{\Delta_+\pi i}{2}}G^{\text{AdS}}_{\Delta_+}(X_{\text{AdS}^+},Y_{\text{AdS}^+})+\,\left(\Delta_+ \to \Delta_-\right),\\
 K^{(0)}_{\Delta,\,
 \pm}(X_{\text{dS}^+},Q_+)&\to c^{\text{dS-AdS}}_{\Delta}e^{\mp \frac{\Delta \pi i}{2}}K^{\text{AdS}}_{\Delta}(X_{\text{AdS}^+},Q_+),\\
 \int_\pm {\rm d}^{d+1} X_{\text{dS}^+} &\to e^{\pm \frac{d \pi i}{2}} \int {\rm d}^{d+1} X_{\text{AdS}^+},
 \end{align}   
\end{subequations}
where
   \begin{equation}\label{dsads2pt}
   c^{\text{dS-AdS}}_{\Delta} = \frac{C^{\text{dS}}_{\Delta}}{C^{\text{AdS}}_{\Delta}}=\frac{1}{2}\csc\left(\tfrac{\pi}{2}\left(d-2\Delta\right)\right),
   \end{equation}
accounts for the change in two-point coefficient between AdS and dS. The upshot is that any given perturbative contribution to a late-time correlator in the Bunch-Davies vacuum can be expressed in terms of corresponding Witten diagrams in EAdS.

\vskip 4pt
In the following we extend the above results to a generic Bogoliubov initial state, focusing first on $\alpha$-vacua where, owing to dS invariance, the statement can be readily formulated in position space. In section \ref{subsec::alphafromBD} we saw that the Schwinger-Keldysh propagators in a generic $\alpha$-vacuum can be expressed in terms of their Bunch-Davies counterparts with points antipodally transformed to the contracting Poincar\'e patch. Using the above analytic continuations, which are valid for both the expanding and contracting patches, two-point functions in a generic $\alpha$-vacuum can also be expressed in terms of corresponding propagators in EAdS but with points also on the lower-sheet of the EAdS hyperboloid. In particular, we have that 
\begin{subequations}
 \begin{align}
\hspace*{-0.5cm} G^{(0)}_{\pm {\hat \pm}}({\bar X}_{\text{dS}^+},{\bar Y}_{\text{dS}^+})&\to c^{\text{dS-AdS}}_{\Delta_+}e^{\mp \frac{\Delta_+\pi i}{2}}e^{{\hat \mp} \frac{\Delta_+\pi i}{2}}G^{\text{AdS}}_{\Delta_+}(\sigma({\bar X}_{\text{AdS}^+},{\bar Y}_{\text{AdS}^+}))+\,\left(\Delta_+ \to \Delta_-\right),\\
 \hspace*{-0.5cm} G^{(0)}_{- +}({\bar X}_{\text{dS}^+},Y_{\text{dS}^+})&\to c^{\text{dS-AdS}}_{\Delta_+}G^{\text{AdS}}_{\Delta_+}(\sigma({\bar X}_{\text{AdS}^+},Y_{\text{AdS}^+})-i\epsilon)+\,\left(\Delta_+ \to \Delta_-\right),\\
\hspace*{-0.5cm}  G^{(0)}_{- +}(X_{\text{dS}^+},{\bar Y}_{\text{dS}^+})&\to c^{\text{dS-AdS}}_{\Delta_+}G^{\text{AdS}}_{\Delta_+}(\sigma(X_{\text{AdS}^+},{\bar Y}_{\text{AdS}^+})+i\epsilon)+\,\left(\Delta_+ \to \Delta_-\right),\\
 \hspace*{-0.5cm}  G^{(0)}_{+ -}({\bar X}_{\text{dS}^+},Y_{\text{dS}^+})&\to c^{\text{dS-AdS}}_{\Delta_+}G^{\text{AdS}}_{\Delta_+}(\sigma({\bar X}_{\text{AdS}^+},Y_{\text{AdS}^+})+i\epsilon)+\,\left(\Delta_+ \to \Delta_-\right),\\
\hspace*{-0.5cm}  G^{(0)}_{+
-}(X_{\text{dS}^+},{\bar Y}_{\text{dS}^+})&\to c^{\text{dS-AdS}}_{\Delta_+}G^{\text{AdS}}_{\Delta_+}(\sigma(X_{\text{AdS}^+},{\bar Y}_{\text{AdS}^+})-i\epsilon)+\,\left(\Delta_+ \to \Delta_-\right),
 \end{align}    
\end{subequations}
and bulk-to-boundary propagators
 \begin{subequations}
\begin{align}\label{analdSbuboT}
K^{(0)}_{\Delta,\,+}({\bar X}_{\text{dS}^+},Q)&\to  c^{\text{dS-AdS}}_{\Delta}\,i^{+ \Delta}\,K^{\text{AdS}}_{\Delta}(s_{\text{AdS}}({\bar X}_{\text{AdS}^+},Q)-i\epsilon),\\
K^{(0)}_{\Delta,\,-}({\bar X}_{\text{dS}^+},Q)&\to  c^{\text{dS-AdS}}_{\Delta}\,i^{- \Delta}\,K^{\text{AdS}}_{\Delta}(s_{\text{AdS}}({\bar X}_{\text{AdS}^+},Q)+i\epsilon),
\end{align}   
\end{subequations}
where ${\bar X}_{\text{(A)dS}^+}=X_{\text{(A)dS}^{-}}$ is the antipodal point to $X_{{\text{(A)dS}}^{+}}$. Combined with the expressions in section \ref{subsec::alphavacua} for propagators in a generic $\alpha$-vacuum in terms of their Bunch-Davies counterparts, this shows that the Feynman rules for the corresponding late-time correlators can be recast in terms of propagators for fields of the same mass in EAdS. Unlike for the Bunch-Davies vacuum however, such propagators have points on the lower sheet of the EAdS hyperboloid. See figure \ref{fig::ds2ads}.

\vskip 4pt
From the above we see that, unlike for the Bunch-Davies vacuum, late-time correlators in a generic dS invariant vacuum do not share the same analytic structure as EAdS boundary correlators encountered in the context of the AdS/CFT correspondence -- where all points are on the upper sheet of the EAdS hyperboloid. This statement is particularly clear-cut in momentum space, where late-time correlators in a generic $\alpha$-vacuum are well known to exhibit singularities for collinear momentum configurations (see section \ref{sec::CWI}), which are the hallmark of antipodal singularities in momentum-space. This can be understood from the expressions \eqref{bubufromBD} and \eqref{bubofromBD} for Schwinger-Keldysh propagators in a generic $\alpha$-vacuum, which are given in terms of those in the Bunch-Davies vacuum with magnitudes of the momenta rotated by $e^{\pm \pi i}$. This converts the ``total/partial energy" singularities of boundary correlators in the Bunch-Davies vacuum \cite{Maldacena:2011nz,Raju:2012zr,Baumann:2021fxj} into collinear singularities. One can then apply the analytic continuation \eqref{poinanal} to obtain analogous expressions in EAdS:\footnote{As we shall see explicitly in the next section, rotations of the internal momenta associated to bulk-to-bulk propagators can be traded with rotations of the external momenta by dilatation symmetry.}
\begin{subequations}\label{bubuan}
\begin{align}\nonumber
 G^{(\alpha)}_{++}(\eta_1,k;\eta_2,k)&\to e^{-\Delta_+ \pi i} c^{\text{dS-AdS}}_{\Delta_+} \left(P^+_{\Delta_+} G^{\text{AdS}}_{\Delta_+}(z_1,k;z_2,k)+e^{-2\nu\pi }P^-_{\Delta_+}  G^{\text{AdS}}_{\Delta_+}(z_1,{\bar k}^+;z_2,{\bar k}^+)\right)\\& \quad + (\Delta_+ \to \Delta_-),\\ \nonumber
G^{(\alpha)}_{--}(\eta_1,k;\eta_2,k)&\to e^{\Delta_+ \pi i}c^{\text{dS-AdS}}_{\Delta_+}\left( M^-_{\Delta_+} G^{\text{AdS}}_{\Delta_+}(z_1,k;z_2,k)+ \, e^{2\nu \pi }  M^+_{\Delta_+}G^{\text{AdS}}_{\Delta_+}(z_1,{\bar k}^-;z_2,{\bar k}^-)\right)\\ & \quad + (\Delta_+ \to \Delta_-), \\ \nonumber
G^{(\alpha)}_{-+}(\eta_1,k;\eta_2,k)& \to \cosh^2\alpha\, c^{\text{dS-AdS}}_{\Delta_+}\, G^{\text{AdS}}_{\Delta_+}(z_1,k;z_2,k)+\sinh^2\alpha \,e^{2\nu \pi}c^{\text{dS-AdS}}_{\Delta_+}G^{\text{AdS}}_{\Delta_+}(z_1,{\bar k}^-;z_2,{\bar k}^+)\\ \nonumber &-\frac{1}{2}\sinh 2\alpha\,e^{\nu \pi } c^{\text{dS-AdS}}_{\Delta_+} [ e^{i\beta}\, G^{\text{AdS}}_{\Delta_+}(z_1,k;z_2,{\bar k}^+)+ e^{-i\beta} \, G^{\text{AdS}}_{\Delta_+}(z_1,{\bar k}^-;z_2,k)]\\ &+ (\Delta_+ \to \Delta_-), \\ \nonumber
G^{(\alpha)}_{+-}(\eta_1,k;\eta_2,k) &\to \cosh^2\alpha \, c^{\text{dS-AdS}}_{\Delta_+} \, G^{\text{AdS}}_{\Delta_+}(z_1,k;z_2,k)+\sinh^2\alpha \,e^{-2\nu \pi } c^{\text{dS-AdS}}_{\Delta_+} G^{\text{AdS}}_{\Delta_+}(z_1,{\bar k}^+;z_2,{\bar k}^-)\\ \nonumber
 &-\frac{1}{2}\sinh 2\alpha\,e^{-\nu\pi}c^{\text{dS-AdS}}_{\Delta_+}[ e^{i\beta}\, G^{\text{AdS}}_{\Delta_+}(z_1,k;z_2,{\bar k}^-)+ e^{-i\beta} \, G^{\text{AdS}}_{\Delta_+}(z_1,{\bar k}^+;z_2,k)]\\ &+ (\Delta_+ \to \Delta_-), 
\end{align}    
\end{subequations}
and
\begin{subequations}\label{buboan}
 \begin{align}
\hspace*{-0.5cm}K^{(\alpha)}_{\Delta,\, +}\left(\eta,k\right) &\to e^{-\frac{\Delta \pi i}{2}} c^{\text{dS-AdS}}_{\Delta} \left(-\eta_0\right)^{\Delta} \left( P^+_{\Delta} K^{\text{AdS}}_{\Delta}\left(z,k\right) + P^-_{\Delta} e^{(2\Delta-d) \pi i}K^{\text{AdS}}_{\Delta}(z,{\bar k}^+)\right),\\
\hspace*{-0.5cm} K^{(\alpha)}_{\Delta,\, -}\left(\eta,k\right) &\to e^{+\frac{\Delta \pi i}{2}}c^{\text{dS-AdS}}_{\Delta}\left(-\eta_0\right)^{\Delta}\left( M^+_{\Delta} e^{-(2\Delta-d) \pi i}K^{\text{AdS}}_{\Delta}(z,{\bar k}^-) + M^-_{\Delta} K^{\text{AdS}}_{\Delta}\left(z,k\right)\right).
\end{align}   
\end{subequations}
In momentum space these relations extend to a generic Bogoliubov initial state by allowing $\alpha$ and $\beta$ to depend on the momentum.

\vskip 4pt
In cosmology, due to translation invariance we are often interested in late-time correlators in momentum space. In the following sections we will apply the above identities in some examples to compute perturbative late-time correlation functions in momentum space by re-writing them in terms of corresponding EAdS Witten diagrams, focusing for ease of presentation on $\alpha$-vacua.

\section{Perturbative Calculations}
\label{sec::PC}

In this section we apply the rules outlined in section \ref{sec::SFEPP} in some examples to re-write perturbative contributions to late-time correlators in terms of corresponding Witten diagrams in EAdS. For simplicity we focus on correlators in $\alpha$-vacua, where $\alpha$ and $\beta$ are momentum independent, though the results extend straightforwardly to generic Bogoliubov initial states.

\subsection{Two-point function}

Let us begin with the late-time free theory two-point function in a generic $\alpha$-vacuum. This can be simply obtained from the late-time limit of the corresponding Wightman function and, in turn, from the Bunch-Davies result via \eqref{wafrombd}:
\begin{align}
  \lim_{\eta_1,\eta_2 \to 0}  \langle \phi(\eta_1)\phi(\eta_2) \rangle^{(\alpha)} &= \lim_{\eta_1,\eta_2 \to 0}G^{(\alpha)}_{W}(\eta_1,\eta_2), \\ \nonumber &= \lim_{\eta_1,\eta_2 \to 0} \left[\cosh^2\alpha \,  \langle \phi(\eta_1)\phi(\eta_2) \rangle^{(0)}+\sinh^2\alpha \, \langle \phi({\bar \eta}^-_1)\phi({\bar \eta}^+_2) \rangle^{(0)}\right.\\\nonumber & \hspace*{-0.5cm} -\left. \frac{1}{2}\sinh 2\alpha \left( e^{i\left(\beta+\frac{\pi d}{2}\right)} \, \langle \phi(\eta_1)\phi({\bar \eta}^+_2) \rangle^{(0)}
  + e^{-i\left(\beta+\frac{\pi d}{2}\right)} \, \langle \phi({\bar \eta}^-_1)\phi(\eta_2) \rangle^{(0)}\right)\right],
\end{align}
where 
\begin{align}\nonumber
    \lim_{\eta_1,\eta_2 \to 0}\langle \phi(\eta_1)\phi(\eta_2) \rangle^{(0)}&=\lim_{\eta_1,\eta_2 \to 0}G^{(0)}_{W}(\eta_1;\eta_2),\\
    &= \lim_{\eta_1,\eta_2 \to 0}\left[G^{(0)}_{\Delta_+,W}(\eta_1;\eta_2)+G^{(0)}_{\Delta_-,W}(\eta_1;\eta_2)\right],
\end{align}
with (see appendix \ref{app::PI})
\begin{equation}
    \lim_{\eta_1,\eta_2 \to 0}G^{(0)}_{\Delta_+,W}(\eta_1,\vec{k};\eta_2,-\vec{k}\,)=\frac{\left(\eta_1 \eta_2\right)^{\frac{d}{2}}}{4\pi}\Gamma\left(\frac{d}{2}-\Delta_+\right)^2\left(\frac{\eta_1\eta_2 k^2}{4}\right)^{\Delta_+-\frac{d}{2}}+\ldots,
\end{equation}
where $\ldots$ are local terms that are analytic in $k$ and do not encode long range interactions.

\vskip 4pt
All together this gives:
\begin{multline}\label{dS2pt}
    \lim_{\eta_1,\eta_2 \to 0}  \langle \phi(\eta_1)\phi(\eta_2) \rangle^{(\alpha)} = \left[\cosh 2\alpha - \sinh 2\alpha \cos\left(\pi \left(\tfrac{\beta}{\pi}+\tfrac{d}{2}-\Delta_+\right)\right)\right]\lim_{\eta_1,\eta_2 \to 0}  G^{(0)}_{\Delta_+,W}(\eta_1;\eta_2)\\
    +\left(\Delta_+ \to \Delta_-\right).
\end{multline}
For a massless scalar (i.e. $\nu=\frac{id}{2}$) in $d=3$ this reduces to
\begin{equation}\label{masslessdS2ptd3}
\lim_{\eta_1,\eta_2 \to 0}  \langle \phi_{\vec{k}}(\eta_1)\phi_{-\vec{k}}(\eta_2) \rangle^{(\alpha)} = \frac{\cosh 2\alpha +\sin \beta \sinh 2\alpha}{2k^3}.
\end{equation}

\subsection{Contact diagrams}

Consider the $n$-point late-time contact diagram generated by the non-derivative interaction
\begin{equation}\label{npointvertex}
    {\cal V}_{12 \ldots n} = - g \phi_1 \phi_2 \ldots \phi_n,
\end{equation}
of scalar fields $\phi_i$, $i=1,2,\ldots,n$ with dS mass $m^2_i = \Delta_i(d-\Delta_i)$. In the Schwinger-Keldysh formalism, in a generic $\alpha$-vacuum this is given by
\begin{equation}
    {}^{(\alpha)}{\cal A}^{{\cal V}_{12 \ldots n}}_{\Delta_1 \Delta_2 \ldots \Delta_n} = {}^{(\alpha,+)}{\cal A}^{{\cal V}_{12 \ldots n}}_{\Delta_1 \Delta_2 \ldots \Delta_n}+{}^{(\alpha,-)}{\cal A}^{{\cal V}_{12 \ldots n}}_{\Delta_1 \Delta_2 \ldots \Delta_n},
\end{equation}
with contributions from the time-ordered ($+$) and anti-time-ordered ($-$) branches:
\begin{equation}
    {}^{(\alpha,\pm)}{\cal A}^{{\cal V}_{1 \ldots n}}_{\Delta_1 \ldots \Delta_n}\left(k_1,k_2,\ldots,k_n\right) = \pm i g\int^{\eta_0}_{-\infty}\frac{{\rm d}\eta}{\left(-\eta\right)^{d+1}}\,K^{(\alpha)}_{\Delta_1,\pm}\left(\eta,k_1\right) \ldots K^{(\alpha)}_{\Delta_n,\pm}\left(\eta,k_n\right),
\end{equation}
where $\eta_0\sim0$. Using the expressions \eqref{bubofromBD} for bulk-to-boundary propagators for generic $\alpha$ in terms of those in the Bunch-Davies vacuum $(\alpha=0)$, each contribution can be re-cast in terms of the corresponding contribution in the Bunch-Davies vacuum with the momenta appropriately rotated: 
\begin{align}
   & {}^{(\alpha,+)}{\cal A}^{{\cal V}_{1 \ldots n}}_{\Delta_1 \ldots \Delta_n}\left(k_1,k_2,\ldots,k_n\right)=P_{\Delta_1}^{+} \ldots P_{\Delta_n}^{+}\,{}^{(0,+)}{\cal A}^{{\cal V}_{1 \ldots n}}_{\Delta_1 \ldots \Delta_n}\left(k_1,k_2,\ldots,k_n\right)\\
    &+\sum\limits_i e^{(2\Delta_i-d) \pi i}P_{\Delta_1}^{+} \ldots P_{\Delta_i}^{-} \ldots P_{\Delta_n}^{+}\,{}^{(0,+)}{\cal A}^{{\cal V}_{1 \ldots n}}_{\Delta_1 \ldots \Delta_n}\left(k_1,\ldots, {\bar k}^+_i,\ldots, k_n\right) \nonumber\\ \nonumber
    &+\sum\limits_{i\,<\,j} e^{(2\Delta_i-d) \pi i}e^{(2\Delta_j-d) \pi i}P_{\Delta_1}^{+}\ldots P_{\Delta_i}^{-}\ldots P_{\Delta_j}^{-} \ldots P_{\Delta_n}^{+}\,{}^{(0,+)}{\cal A}^{{\cal V}_{1 \ldots n}}_{\Delta_1 \ldots \Delta_n}\left(k_1, \ldots, {\bar k}^+_i,\ldots,  {\bar k}^+_j,\ldots,k_n\right) \nonumber\\ \nonumber &\vdots\\
    &+e^{(2\Delta_1-d) \pi i}\ldots e^{(2\Delta_n-d) \pi i}P_{\Delta_1}^{-} \ldots P_{\Delta_n}^{-}\,{}^{(0,+)}{\cal A}^{{\cal V}_{1 \ldots n}}_{\Delta_1 \ldots \Delta_n}\left({\bar k}^+_1,\ldots,{\bar k}^+_n\right), \nonumber
\end{align}
and 
\begin{align}
   & {}^{(\alpha,-)}{\cal A}^{{\cal V}_{1 \ldots n}}_{\Delta_1 \ldots \Delta_n}\left(k_1,k_2,\ldots,k_n\right)=M_{\Delta_1}^{-} \ldots M_{\Delta_n}^{-}\,{}^{(0,-)}{\cal A}^{{\cal V}_{1 \ldots n}}_{\Delta_1 \ldots \Delta_n}\left(k_1,k_2,\ldots,k_n\right)\\
    &+\sum\limits_i e^{-(2\Delta_i-d) \pi i}M_{\Delta_1}^{-} \ldots M_{\Delta_i}^{+} \ldots M_{\Delta_n}^{-}\,{}^{(0,-)}{\cal A}^{{\cal V}_{1 \ldots n}}_{\Delta_1 \ldots \Delta_n}\left(k_1,\ldots, {\bar k}^-_i,\ldots, k_n\right) \nonumber\\ \nonumber
    &+\sum\limits_{i\,<\,j} e^{-(2\Delta_i-d) \pi i}e^{-(2\Delta_j-d) \pi i}M_{\Delta_1}^{-}\ldots M_{\Delta_i}^{+}\ldots M_{\Delta_j}^{+} \ldots M_{\Delta_n}^{-}\,{}^{(0,-)}{\cal A}^{{\cal V}_{1 \ldots n}}_{\Delta_1 \ldots \Delta_n}\left(k_1, \ldots, {\bar k}^-_i,\ldots,  {\bar k}^-_j,\ldots,k_n\right) \nonumber\\ \nonumber &\vdots\\
    &+e^{-(2\Delta_1-d) \pi i}\ldots e^{-(2\Delta_n-d) \pi i}M_{\Delta_1}^{+} \ldots M_{\Delta_n}^{+}\,{}^{(0,-)}{\cal A}^{{\cal V}_{1 \ldots n}}_{\Delta_1 \ldots \Delta_n}\left({\bar k}^-_1,\ldots,{\bar k}^-_n\right). \nonumber
\end{align}
These, in turn, can be expressed in terms of corresponding Witten diagrams in EAdS using the analytic continuations \eqref{buboan}, where \cite{Sleight:2019mgd}: 
\begin{multline}\label{pmcontads}
   {}^{(0,\pm)}{\cal A}^{{\cal V}_{1 \ldots n}}_{\Delta_1 \ldots \Delta_n}\left(k_1,k_2,\ldots,k_n\right) \\ = \pm i\, e^{\mp \frac{i \pi}{2}\left(\frac{(n-2)d}{2}+ i (\nu_1+\ldots+\nu_n)\right)}\left(\prod\limits^n _{i=1} c^{\text{dS-AdS}}_{\Delta_i}\left(-\eta_0\right)^{\Delta_i}\right){\cal A}^{\text{AdS}}_{\Delta_1 \ldots \Delta_n}\left(k_1,\ldots,k_n\right),
\end{multline}
with
\begin{equation}\label{nK}
    {\cal A}^{\text{AdS}}_{\Delta_1 \ldots \Delta_n}\left(k_1,\ldots,k_n\right)= -g \int^{\infty}_{0}\frac{{\rm d}z}{z^{d+1}}\,K^{\text{AdS}}_{\Delta_1}\left(z,k_1\right) \ldots K^{\text{AdS}}_{\Delta_n}\left(z,k_n\right).
\end{equation}
As explained in section \ref{subsec::toEAdS}, rotations in the momenta are equivalent to the antipodal transformation of the corresponding bulk point. In particular, 
\begin{multline}
{\cal A}^{\text{AdS}}_{\Delta_1 \ldots \Delta_n}\left(k_1,\ldots,{\bar k}^\pm_i,\ldots, k_n\right)\\= - g (e^{\mp \pi i})^{\Delta_i-d}\int^{\infty}_{0}\frac{{\rm d}z}{z^{d+1}}\,K^{\text{AdS}}_{\Delta_1}\left(z,k_1\right) \ldots K^{\text{AdS}}_{\Delta_i}\left({\bar z}^\pm ,k_1\right) \ldots K^{\text{AdS}}_{\Delta_n}\left(z,k_n\right).
\end{multline}

\subsubsection{Three-point}

Let us examine the case $n=3$ in more detail. In this case the dilatation Ward identity implies that (see appendix \ref{app::CI}):
\begin{equation}\label{3ptmomflip}
   {\cal A}^{\text{AdS}}_{\Delta_1 \Delta_2 \Delta_3}\left(k_1e^{\mp i\pi},k_2e^{\mp i\pi},k_3\right)=\left(e^{\mp \pi i}\right)^{ \Delta_1+\Delta_2+\Delta_3-2d}{\cal A}^{\text{AdS}}_{\Delta_1 \Delta_2 \Delta_3}\left(k_1,k_2,k_3e^{\pm i\pi}\right),
\end{equation}
which gives rise to the following compact expression:
\begin{align}\label{3pt}
 & {}^{(\alpha)}{\cal A}^{{\cal V}_{123}}_{\Delta_1 \Delta_2 \Delta_3}\left(k_1,k_2,k_3\right)\\ \nonumber &= i \left(\prod\limits^3 _{i=1} c^{\text{dS-AdS}}_{\Delta_i}\left(-\eta_0\right)^{\Delta_i}\right) \sum_{\pm} e^{\mp \frac{i \pi}{2}\left(\frac{d}{2}+ i (\nu_1+\nu_2+\nu_3)\right)}\left\{\left(P^{\nu_1}_{\pm} P^{\nu_2}_{\pm} P^{\nu_3}_{\pm}-M^{\nu_1}_{\pm}M^{\nu_2}_{\pm}M^{\nu_3}_{\pm}\right){\cal A}^{\text{AdS}}_{\Delta_1 \Delta_2 \Delta_3}\left(k_1,k_2,k_3\right)\right. \\ \nonumber
    &\hspace*{4.9cm}+  \left(P^{\nu_1}_{\mp} P^{\nu_2}_{\pm} P^{\nu_3}_{\pm}-M^{\nu_1}_{\mp}M^{\nu_2}_{\pm}M^{\nu_3}_{\pm}\right)e^{\mp 2 \pi \nu_1}{\cal A}^{\text{AdS}}_{\Delta_1 \Delta_2 \Delta_3}(e^{\mp i \pi}k_1,k_2,k_3)\\ \nonumber
    &\hspace*{4.9cm}+ \left(P^{\nu_1}_{\pm} P^{\nu_2}_{\mp} P^{\nu_3}_{\pm}-M^{\nu_1}_{\pm}M^{\nu_2}_{\mp}M^{\nu_3}_{\pm}\right)e^{\mp 2 \pi \nu_2}{\cal A}^{\text{AdS}}_{\Delta_1 \Delta_2 \Delta_3}(k_1,e^{\mp i \pi}k_2,k_3)\\ \nonumber
    &\hspace*{4.9cm}\left.+ \left(P^{\nu_1}_{\pm} P^{\nu_2}_{\pm} P^{\nu_3}_{\mp}-M^{\nu_1}_{\pm}M^{\nu_2}_{\pm}M^{\nu_3}_{\mp}\right)e^{\mp 2 \pi \nu_3}{\cal A}^{\text{AdS}}_{\Delta_1 \Delta_2 \Delta_3}(k_1,k_2,e^{\mp i \pi}k_3)\right\}.
\end{align}
It is also convenient to express the contact diagram in the following form, which makes manifest that one recovers the known Bunch-Davies result upon setting $\alpha = 0$: 
\begin{multline}\label{3ptcontactdS}
    {}^{(\alpha)}{\cal A}^{{\cal V}_{123}}_{\Delta_1 \Delta_2 \Delta_3}\left(k_1,k_2,k_3\right) =  i\left(\prod\limits^3 _{i=1} c^{\text{dS-AdS}}_{\Delta_i}\left(-\eta_0\right)^{\Delta_i}\right)\sum_{\pm}e^{\mp \frac{i \pi}{4}(d+2 i (\nu_1+\nu_2+\nu_3))}  \\ \times \left\{  {}^{\left(0\right)}C^\pm_{\Delta_1\Delta_2\Delta_3}   {\cal A}^{\text{AdS}}_{\Delta_1 \Delta_2 \Delta_3}\left(k_1,k_2,k_3\right) 
    + \sinh (2 \alpha )  {}^{\left(\alpha\right)}C^\pm_{\Delta_1\Delta_2\Delta_3}
     \left(  e^{\mp \pi  \nu_1} {\cal A}^{\text{AdS}}_{\Delta_1 \Delta_2 \Delta_3}\left(e^{\mp i \pi}k_1,k_2,k_3\right)\right. \right. \\ \left. \left.+e^{\mp \pi  \nu_2} {\cal A}^{\text{AdS}}_{\Delta_1 \Delta_2 \Delta_3}\left(k_1,e^{\mp i \pi}k_2,k_3\right)  +e^{\mp \pi  \nu_3}{\cal A}^{\text{AdS}}_{\Delta_1 \Delta_2 \Delta_3}\left(k_1,k_2,e^{\mp i \pi}k_3\right) \right)\right\},
\end{multline}
where 
\begin{subequations}\label{Ccoeffs3pt}
 \begin{multline}
   {}^{\left(0\right)}C^\pm_{\Delta_1\Delta_2\Delta_3} = \frac{1}{8}\left[ \pm (3 \cosh (4 \alpha )+5) \mp 2 e^{\mp i \beta } \sinh (4 \alpha ) \left(e^{\mp \pi  \nu_1}+e^{\mp \pi  \nu_2}+e^{\mp \pi  \nu_3}\right) \right. \\   \left. \pm 2 e^{\mp 2i \beta } \sinh ^2(2 \alpha ) \left(e^{\mp \pi  (\nu_1+\nu_2)}+e^{\mp \pi  (\nu_1+\nu_3)}+e^{\mp \pi  (\nu_2+\nu_3)}\right)\right],
\end{multline}
\begin{multline}
   {}^{\left(\alpha\right)}C^\pm_{\Delta_1\Delta_2\Delta_3} = \frac{1}{4}\left[\mp 2 e^{\pm i \beta } \cosh (2 \alpha ) e^{\pm \pi  (\nu_1+\nu_2+\nu_3)} \right. \\ \left.\pm \sinh (2 \alpha ) \left(-e^{\mp 2 i \beta }+e^{\pm \pi  (\nu_1+\nu_2)}+e^{\pm \pi  (\nu_1+\nu_3)}+e^{\pm \pi  (\nu_2+\nu_3)}\right)\right].   
\end{multline}   
\end{subequations}
For $\alpha = 0$ the terms proportional to $\sinh (2 \alpha )$ vanish and ${}^{\left(0\right)}C^\pm_{\Delta_1\Delta_2\Delta_3} = \pm 1$, so that 
\begin{multline}\label{BD3pt}
     {}^{(0)}{\cal A}^{{\cal V}_{123}}_{\Delta_1 \Delta_2 \Delta_3} = 2 \left(\prod\limits^3 _{i=1} c^{\text{dS-AdS}}_{\Delta_i}\left(-\eta_0\right)^{\Delta_i}\right)\, \text{sin}\left(\left(\frac{d}{2}+i\left(\nu_1+\nu_2+\nu_3\right)\right)\frac{\pi}{2}\right) \\ \times {\cal A}^{\text{AdS}}_{\Delta_1 \Delta_2 \Delta_3}\left(k_1,k_2,k_3\right),
\end{multline}
which recovers the result of \cite{Sleight:2019mgd} for contact diagrams in the Bunch-Davies vacuum. 

\vskip 4pt 
For generic scaling dimensions $\Delta_i$ on the Principal Series, conformal three-point functions in Fourier space are given by Appell's $F_4$ function \cite{Bzowski:2013sza}, which is a generalised hypergeometric function of two variables and admits the (``triple-$K$") integral representation \eqref{3K}. In Mellin space, this takes the form \eqref{MBcontact}. Away from the Principal Series, for certain values of the scaling dimensions this expression simplifies and in some cases there are IR divergences. We will see examples of these in the following.

\vskip 4pt
In the following examples we will set the parameter $\beta=0$.

\paragraph{Three massless scalars in $d=3$.} In this case we have $\nu_i = \frac{3i}{2}$ i.e. $\Delta_i=0$: 
\begin{align}\label{massless3pt}
 & \frac{ {}^{(\alpha)}{\cal A}^{{\cal V}_{123}}_{000}\left(k_1,k_2,k_3\right)}{\left(c^{\text{dS-AdS}}_{0}\right)^3} = 2{\cal A}^{\text{AdS}}_{000}\left(k_1,k_2,k_3\right)-\text{sinh}\left(2\alpha\right) \sum\limits_\pm  \left(\text{sinh}\left(2\alpha\right) \pm \frac{i}{2} \text{cosh}\left(2\alpha\right)\right) \\ & \hspace*{2cm}\times \left( {\cal A}^{\text{AdS}}_{000}\left(e^{\mp i \pi}k_1,k_2,k_3\right)+ {\cal A}^{\text{AdS}}_{000}\left(k_1,e^{\mp i \pi}k_2,k_3\right)+{\cal A}^{\text{AdS}}_{000}\left(k_1,k_2,e^{\mp i \pi}k_3\right)\right).\nonumber
\end{align}
The corresponding EAdS contact diagram is given by
\begin{multline}
    {\cal A}^{\text{AdS}}_{000}\left(k_1,k_2,k_3\right) = \frac{1}{3 k^3_1k^3_2k^3_3}\left[2 k^2_1 (k_2+k_3)+2 k_1 \left(k_2^2-k_2 k_3+k_3^2\right)+2 k_2 k_3 (k_2+k_3)\right. \\ \left. +\frac{4 \left(k_1^3+k_2^3+k_3^3\right)}{d-3}-\frac{1}{3} (6 \gamma -11) \left(k_1^3+k_2^3+k_3^3\right) \right. \\ \left. - \left(k_1^3+k_2^3+k_3^3\right) (2 \log (k_1+k_2+k_3)+1)\right],
\end{multline}
which, for example, can be obtained from the Mellin-Barnes representation \eqref{MBcontact} of three-point contact diagrams in dimensional regularisation (see appendix \ref{app::Mellin}). Note that there is a local IR divergence for $d=3$ which can be cancelled in the in-in formalism by adding local counterterms at the future boundary of dS \cite{Bzowski:2023nef}.

\vskip 4pt
The expression \eqref{massless3pt} agrees with the result originally obtained in \cite{Shukla:2016bnu}, up to local terms.

\paragraph{Two conformally coupled and one general scalar.}

In this case we have $\nu_1=\nu_2 = \frac{i}{2}$ and generic $\Delta$:
\begin{align}\label{ccm3pt}
    \frac{{}^{(\alpha)}{\cal A}^{{\cal V}_{123}}_{\frac{d-1}{2}\frac{d-1}{2}\Delta}\left(k_1,k_2,k_3\right)}{\left(\prod\limits^3 _{i=1} c^{\text{dS-AdS}}_{\Delta_i}\left(-\eta_0\right)^{\Delta_i}\right)} &=  \sum_{\pm}  {}^{\left(0\right)}C^\pm_{\frac{d-1}{2}\frac{d-1}{2}\Delta}  e^{\mp \frac{i\pi}{4}  (d-2 i \nu+2 )} {\cal A}^{\text{AdS}}_{\frac{d-1}{2}\frac{d-1}{2}\Delta}\left(k_1,k_2,k_3\right) \\ \nonumber
    & \hspace*{-0.5cm} + \sinh (2 \alpha ) \sum_{\pm} {}^{\left(\alpha\right)}C^\pm_{\frac{d-1}{2}\frac{d-1}{2}\Delta}
     \left[  e^{\mp \frac{i\pi}{4} \left(d-2i \nu+4\right)}{\cal A}^{\text{AdS}}_{\frac{d-1}{2}\frac{d-1}{2}\Delta}\left(e^{\mp i \pi}k_1,k_2,k_3\right)\right.\\ \nonumber & \hspace*{4cm} \left. +e^{\mp \frac{i \pi}{4}   (d-2 i \nu +4)}{\cal A}^{\text{AdS}}_{\frac{d-1}{2}\frac{d-1}{2}\Delta}\left(k_1,e^{\mp i \pi}k_2,k_3\right) \right. \\ \nonumber
   & \left. \hspace*{4cm} +e^{\mp \frac{i \pi}{4} (d-6 i \nu +2)} {\cal A}^{\text{AdS}}_{\frac{d-1}{2}\frac{d-1}{2}\Delta}\left(k_1,k_2,e^{\mp i \pi}k_3\right) \right],
\end{align}
with 
\begin{align}\nonumber
   {}^{\left(0\right)}C^\pm_{\frac{d-1}{2}\frac{d-1}{2}\Delta} &= \frac{1}{8} \left[\mp i (3 \cosh (4 \alpha )+5)\left(\mp 4 e^{\mp \pi  \nu }+2 i\right) \sinh ^2(2 \alpha )+ \left(4\pm2 i e^{\mp \pi  \nu }\right) \sinh (4 \alpha )\right],\\
   {}^{\left(\alpha\right)}C^\pm_{\frac{d-1}{2}\frac{d-1}{2}\Delta} &= \pm \frac{i}{4}\left[2  \cosh (2 \alpha ) e^{\pm \pi  (i+\nu)} + 2 \sinh (2 \alpha ) \left(1\mp i e^{\pm \pi  \nu}\right)\right].
\end{align}
In this case the three-point conformal structure reduces to a Gauss hypergeometric function (see e.g. section 3.3 of \cite{Sleight:2019mgd}):
\begin{multline}\label{ccmads}
   {\cal A}^{\text{AdS}}_{\frac{d-1}{2}\frac{d-1}{2}\Delta}\left(k_1,k_2,k_3\right) = \pi^{3/2} \frac{2^{4-d} }{k_1 k_2}\,\left(\frac{k_3}{2}\right)^{i \nu-\frac{d}{2}+1}\,\frac{\Gamma \left(\tfrac{d}{2}-i \nu-1\right) \Gamma \left(\tfrac{d}{2}+i \nu-1\right)}{\Gamma\left(\tfrac{d-1}{2}\right)}\\ \, \times {}_2F_1\left(\begin{matrix}\tfrac{d}{2}-1 -i\nu,\tfrac{d}{2}-1+i \nu\\\tfrac{d-1}{2}\end{matrix};\frac{k_3-k_1-k_2}{2k_3}\right).
\end{multline}

\paragraph{OPE limit.} It is interesting to use the above examples to study the operator product expansion (OPE) limit $k_3\to0$ of correlators for generic $\alpha$. For a given term in the three-point function \eqref{ccm3pt}, the argument of the Gauss hypergeometric function \eqref{ccmads} takes the form\footnote{Note that by momentum conservation we have $k^2_2 = k^2_1 +k^2_3+2k_1k_3  \cos\,\theta_{k_1,k_3}$.}
\begin{align}
    z &= \frac{k_3-e^{(a-c)\pi i}k_1 + e^{(b-c)\pi i}k_2 }{2 k_3},\,\\&\sim - \frac{1}{2} \left(e^{(a-c)\pi i}+e^{(b-c)\pi i}\right)\frac{k_1}{k_3}+\frac{1}{2}-\frac{1}{2}e^{(b-c)\pi i} \cos\,\theta_{k_1,k_3}+O\left(\frac{k_3}{k_1}\right),
\end{align}
where $a,\,b,\,c \in \left\{0,\pm 1\right\}$ and $\theta_{k_1,k_3}$ is the angle between vectors $\vec{k}_1$ and $\vec{k}_3$. In the Bunch-Davies vacuum (i.e. $a=b=c=0$) the $k_1/k_3$ term is leading, so that 
\begin{equation}
     {}_2F_1\left(\begin{matrix}\tfrac{d}{2}-1 -i\nu,\tfrac{d}{2}-1+i \nu\\\tfrac{d-1}{2}\end{matrix};\frac{k_3-k_1-k_2}{2k_3}\right) \sim \left(\frac{k_1}{k_3}\right)^{-\frac{d}{2}-i \nu +1},
\end{equation}
and 
\begin{equation}\label{ccmbcope}
    {}^{(0)}{\cal A}^{{\cal V}_{123}}_{\frac{d-1}{2}\frac{d-1}{2}\Delta}\left(k_1,k_2,k_3 \to 0\right) \sim \left(\frac{k_3}{k_1}\right)^{2i \nu},
\end{equation}
where we stripped off the overall power of $k_1$ carrying the dimension of the three-point function. In a generic $\alpha$-vacuum instead, for certain values of $a, b, c$ the coefficient of this term vanishes. For such values of $a, b, c$ we have
\begin{multline}
     {\cal A}^{\text{AdS}}_{\frac{d-1}{2}\frac{d-1}{2}\Delta}\left(e^{a \pi i}k_1,e^{b \pi i}k_2,e^{c \pi i}k_3\right) \\ \sim \left(\frac{k_3}{k_1}\right)^{i \nu -\frac{d}{2}+1}{}_2F_1\left(\begin{matrix}\tfrac{d}{2}-1 -i\nu,\tfrac{d}{2}-1+i \nu\\\tfrac{d-1}{2}\end{matrix};\frac{1-e^{(b-c)\pi i} \cos\,\theta_{k_1,k_3}}{2}\right), 
\end{multline}
which, for massive particles (i.e. $\nu \in \mathbb{R}$), dominates over the Bunch-Davies contribution \eqref{ccmbcope} for $d>2$. Note that this develops a branch cut for $\cos\,\theta_{k_1,k_3}=\pm 1$ i.e. when $\vec{k_1}$ and $\vec{k_3}$ are collinear, as expected.

\vskip 4pt
Let us note that this behaviour of three-point correlators for generic $\alpha$ has been interpreted in the literature as being ``inconsistent" with the OPE limit. This however assumes that the solution to the position space conformal Ward identity is unique (up to contact terms). In fact, as we shall see in section \ref{sec::CWI}, the space of solutions to the position space conformal Ward identities must be enlarged to account for those with collinear singularities in momentum space and a correct OPE limit in position space should take the presence of these additional solutions into account.

\subsubsection{Four-point}

One can obtain similar expressions for higher-point contact diagrams. To demonstrate, for $n=4$ we have:
\begin{align}\nonumber
 & \frac{{}^{(\alpha)}{\cal A}^{{\cal V}_{1234}}_{\Delta_1 \Delta_2 \Delta_3 \Delta_4}\left(k_1,k_2,k_3,k_4\right)}{i\left(\prod\limits^4 _{i=1} c^{\text{dS-AdS}}_{\Delta_i}\left(-\eta_0\right)^{\Delta_i}\right)}  = \sum_{\pm}  {}^{\left(0\right)}e^{\mp \frac{i \pi}{2}\left(d+ i (\nu_1+\nu_2+\nu_3+\nu_4)\right)}\left\{C^\pm_{\Delta_1 \Delta_2 \Delta_3 \Delta_4}{\cal A}^{\text{AdS}}_{\Delta_1\Delta_2\Delta_3\Delta_4}\left(k_1,k_2,k_3,k_4\right) \right. \\ \nonumber
    & + \sinh (2 \alpha ) {}^{\left(\alpha, 1\right)}C^\pm_{\Delta_1 \Delta_2 \Delta_3 \Delta_4}\left[e^{\mp \pi \nu_1}{\cal A}^{\text{AdS}}_{\Delta_1 \Delta_2 \Delta_3 \Delta_4}\left(e^{\mp \pi i}k_1,k_2,k_3,k_4\right)+\ldots\right]\\ \nonumber
    &+ \sinh^2 (2 \alpha )  {}^{\left(\alpha, 2\right)}C^\pm_{\Delta_1 \Delta_2 \Delta_3 \Delta_4}\left[e^{\mp \pi\left( \nu_1+\nu_2\right)}{\cal A}^{\text{AdS}}_{\Delta_1 \Delta_2 \Delta_3 \Delta_4}\left(e^{\mp \pi i}k_1,e^{\mp \pi i}k_2,k_3,k_4\right)\right. \\ &\left. \left. +e^{\mp \pi\left( \nu_1+\nu_3\right)}{\cal A}^{\text{AdS}}_{\Delta_1 \Delta_2 \Delta_3 \Delta_4}\left(e^{\mp \pi i}k_1,k_2,e^{\mp \pi i}k_3,k_4\right)+e^{\mp \pi\left( \nu_1+\nu_4\right)}{\cal A}^{\text{AdS}}_{\Delta_1 \Delta_2 \Delta_3 \Delta_4}\left(e^{\mp \pi i}k_1,k_2,k_3,e^{\mp \pi i}k_4\right)\right] \right\},
\end{align}
where
\begin{multline}
{}^{\left(0\right)}C^\pm_{\Delta_1 \Delta_2 \Delta_3 \Delta_4}= \pm \frac{1}{32} \left[4 (7 \cosh (2 \alpha )+\cosh (6 \alpha )) \right. \\ \left.-4 e^{\mp 3 i \beta } \sinh ^3(2 \alpha ) \left(e^{\mp \pi  (\nu_1+\nu_2+\nu_3)}+e^{\mp \pi  (\nu_1+\nu_2+\nu_4)}+e^{\mp \pi  (\nu_1+\nu_3+\nu_4)}+e^{\mp \pi  (\nu_2+\nu_3+\nu_4)}\right)\right. \\ \left.+4 e^{\mp 2 i \beta } \sinh (4 \alpha ) \sinh (2 \alpha ) \left(e^{\mp \pi  (\nu_1+\nu_2)}+e^{\mp \pi  (\nu_1+\nu_3)}+e^{\mp \pi  (\nu_1+\nu_4)}+e^{-\pi  (\nu_2+\nu_3)}+e^{\mp \pi  (\nu_2+\nu_4)}\right. \right. \\\left. \left.+e^{\mp \pi  (\nu_3+\nu_4)}\right)-e^{\mp i \beta } (7 \sinh (2 \alpha )+3 \sinh (6 \alpha )) \left(e^{\mp \pi  \nu_1}+e^{\mp \pi  \nu_2}+e^{\mp \pi  \nu_3}+e^{\mp \pi  \nu_4}\right)\right],
\end{multline}
\begin{multline}
    {}^{\left(\alpha, 1\right)}C^\pm_{\Delta_1 \Delta_2 \Delta_3 \Delta_4} = \pm \frac{1}{16} \left[2 e^{\mp 3 i \beta } e^{\mp \pi  (\nu_1+\nu_2+\nu_3+\nu_4)}\sinh ^2(2 \alpha ) -e^{\pm i \beta } (3 \cosh (4 \alpha )+5)\right.\\\left.-2 e^{\mp i \beta } \sinh ^2(2 \alpha ) \left(e^{\mp \pi  (\nu_1+\nu_2)}+e^{\mp \pi  (\nu_1+\nu_3)}+e^{\mp \pi  (\nu_1+\nu_4)}+e^{\mp \pi  (\nu_2+\nu_3)}+e^{\mp \pi  (\nu_2+\nu_4)}+e^{\mp \pi  (\nu_3+\nu_4)}\right)\right.\\\left.+2 \sinh (4 \alpha ) \left(e^{\mp \pi  \nu_1}+e^{\mp \pi  \nu_2}+e^{\mp \pi  \nu_3}+e^{\mp \pi  \nu_4}\right)\right],
\end{multline}
\begin{multline}
    {}^{\left(\alpha, 2\right)}C^\pm_{\Delta_1 \Delta_2 \Delta_3 \Delta_4}=\pm \frac{1}{8} \left[-e^{\pm i \beta } \sinh (2 \alpha ) \left(e^{\mp \pi  \nu_1}+e^{\mp \pi  \nu_2}+e^{\mp \pi  \nu_3}+e^{\mp \pi  \nu_4}\right)\right.\\\left.+e^{\mp i \beta } \sinh (2 \alpha ) \left(e^{\mp \pi  (\nu_1+\nu_2+\nu_3)}+e^{\mp \pi  (\nu_1+\nu_2+\nu_4)}+e^{\mp \pi  (\nu_1+\nu_3+\nu_4)}+e^{\mp \pi  (\nu_2+\nu_3+\nu_4)}\right)\right.\\\left.-2 e^{\mp 2 i \beta } \cosh (2 \alpha ) e^{\mp \pi  (\nu_1+\nu_2+\nu_3+\nu_4)}+2 e^{\pm 2 i \beta } \cosh (2 \alpha )\right],
\end{multline}
extending the coefficients \eqref{Ccoeffs3pt} for generic scalars to $n=4$.

\subsection{Exchange diagrams}

In the same way one can recast four-point exchange diagrams in a generic $\alpha$-vacuum in terms of corresponding exchanges in Euclidean AdS space with the momenta appropriately rotated. In this case one also makes use of the identities \eqref{bubufromBD} for the bulk-to-bulk propagators in the Schwinger-Keldysh formalism which, under the analytic continuations in section \ref{subsec::toEAdS}, are expressed in terms of corresponding bulk-to-bulk propagators in EAdS with rotations of the exchanged momentum.

\vskip 4pt
In fact, due to symmetry under dilatations, it is possible to trade such rotations of the exchanged momenta for rotations of the momentum of the external fields. Consider the exchange diagram mediated by the cubic vertices:
\begin{equation}\label{cubicV}
{\cal V}_{12\phi} = g_{12} \phi_1 \phi_2 \phi, \qquad {\cal V}_{34\phi} = g_{34} \phi_3 \phi_4 \phi.
\end{equation}
The contribution to the $s$-channel exchange from the $++$ branch of the in-in contour (i.e. both bulk points time-ordered) is
\begin{multline}
 {}^{(\alpha,++)}{\cal A}^{{\cal V}_{12\phi}{\cal V}_{34\phi}}_{\Delta_1\Delta_2 \Delta_3 \Delta_4}\left(k_1,k_2,k_3,k_4\right) = \left(+i g_{12}\right)\left(+i g_{34}\right) \int^0_{-\infty} \frac{{\rm d}\eta_1}{\left(-\eta_1\right)^{d+1}}\int^0_{-\infty} \frac{{\rm d}\eta_2}{\left(-\eta_2\right)^{d+1}}\\ 
 \times K^{(\alpha)}_{\Delta_1,+}\left(\eta_1,k_1\right)K^{(\alpha)}_{\Delta_2,+}\left(\eta_1,k_2\right)K^{(\alpha)}_{\Delta_3,+}\left(\eta_2,k_3\right)K^{(\alpha)}_{\Delta_4,+}\left(\eta_2,k_4\right)\\
 \times \left[P^+_{\Delta_+} G^{(0)}_{\Delta_+,++}(\eta_1,k;\eta_2,k)+e^{-2\nu\pi }P^-_{\Delta_+}  G^{(0)}_{\Delta_+,++}(\eta_1,{\bar k}^+;\eta_2,{\bar k}^+)+ \left(\Delta_+ \to \Delta_-\right)\right] ,
\end{multline}
where we inserted the expression \eqref{bubufromBD} for the $\alpha$-vacuum bulk-to-bulk propagator in terms of its counterpart in the Bunch-Davies vacuum. Using constraints from dilatation symmetry this can be written in the form (see appendix \ref{app::CI})
\begin{multline}
 {}^{(\alpha,++)}{\cal A}^{{\cal V}_{12\phi}{\cal V}_{34\phi}}_{\Delta_1\Delta_2 \Delta_3 \Delta_4}\left(k_1,k_2,k_3,k_4\right) = P^+_{\Delta_+}  {}^{(\alpha,++)}{\cal A}^{{\cal V}_{12\phi}{\cal V}_{34\phi}}_{\Delta_1\Delta_2|\Delta_+| \Delta_3 \Delta_4}\left(k_1,k_2,k_3,k_4\right)\\+P^-_{\Delta_+}\left(e^{+i\pi}\right)^{2\Delta_+-i(\nu_1+i\nu_2+i\nu_3+i\nu_4)}  {}^{(\alpha,++)}{\cal A}^{{\cal V}_{12\phi}{\cal V}_{34\phi}}_{\Delta_1\Delta_2|\Delta_+| \Delta_3 \Delta_4}\left({\bar k}^-_1,{\bar k}^-_2,{\bar k}^-_3,{\bar k}^-_4\right)\\+ \left(\Delta_+ \to \Delta_-\right),
\end{multline}
with only the external momenta $k_i$ rotated, where
\begin{multline}
{}^{(\alpha,++)}{\cal A}^{{\cal V}_{12\phi}{\cal V}_{34\phi}}_{\Delta_1\Delta_2|\Delta_+| \Delta_3 \Delta_4}\left(k_1,k_2,k_3,k_4\right)=\left(+i g_{12}\right)\left(+i g_{34}\right) \int^0_{-\infty} \frac{{\rm d}\eta_1}{\left(-\eta_1\right)^{d+1}}\int^0_{-\infty} \frac{{\rm d}\eta_2}{\left(-\eta_2\right)^{d+1}}\\ 
 \times K^{(\alpha)}_{\Delta_1,+}\left(\eta_1,k_1\right)K^{(\alpha)}_{\Delta_2,+}\left(\eta_1,k_2\right)G^{(0)}_{\Delta_+,++}(\eta_1,k;\eta_2,k)
K^{(\alpha)}_{\Delta_3,+}\left(\eta_2,k_3\right)K^{(\alpha)}_{\Delta_4,+}\left(\eta_2,k_4\right).
\end{multline}
Applying the identities \eqref{bubofromBD} for the bulk-to-boundary propagators as well, this can be expressed in terms of exchanges in the Bunch-Davies vacuum:
\begin{multline}
    {}^{(\alpha,++)}{\cal A}^{{\cal V}_{12\phi}{\cal V}_{34\phi}}_{\Delta_1 \Delta_2|\Delta_+| \Delta_3 \Delta_4}\left(k_1,k_2,k_3,k_4\right)=P_{\Delta_1}^{+}P_{\Delta_2}^{+}P_{\Delta_3}^{+} P_{\Delta_4}^{+}\,{}^{(0,++)}{\cal A}^{{\cal V}_{12\phi}{\cal V}_{34\phi}}_{\Delta_1 \Delta_2 |\Delta_+| \Delta_3 \Delta_4}\left(k_1,k_2,k_3,k_4\right)\\
    +\sum\limits_i e^{(2\Delta_i-d) \pi i}P_{\Delta_1}^{+} \ldots P_{\Delta_i}^{-} \ldots P_{\Delta_4}^{+}\,{}^{(0,++)}{\cal A}^{{\cal V}_{12\phi}{\cal V}_{34\phi}}_{\Delta_1 \Delta_2 |\Delta_+| \Delta_3 \Delta_4}\left(k_1,\ldots, {\bar k}^+_i,\ldots, k_4\right) \\ 
    +\sum\limits_{i\,<\,j} e^{(2\Delta_i-d) \pi i}e^{(2\Delta_j-d) \pi i}P_{\Delta_1}^{+}\ldots P_{\Delta_i}^{-}\ldots P_{\Delta_j}^{-} \ldots P_{\Delta_4}^{+}\,{}^{(0,++)}{\cal A}^{{\cal V}_{12\phi}{\cal V}_{34\phi}}_{\Delta_1 \Delta_2 |\Delta_+| \Delta_3 \Delta_4}\left(k_1, \ldots, {\bar k}^+_i,\ldots,  {\bar k}^+_j,\ldots,k_4\right) \\
    + e^{\sum\limits_i(2\Delta_i-d) \pi i} \sum\limits_i e^{-(2\Delta_i-d) \pi i} P_{\Delta_1}^{-} \ldots P_{\Delta_i}^{+} \ldots P_{\Delta_4}^{-} {}^{(0,++)}{\cal A}^{{\cal V}_{12\phi}{\cal V}_{34\phi}}_{\Delta_1 \Delta_2 |\Delta_+| \Delta_3 \Delta_4}\left({\bar k}^+_1, \ldots, k_i,\ldots,{\bar k}^+_4\right)\\
    +\left(\prod^4_{i=1}e^{(2\Delta_i-d) \pi i}P_{\Delta_i}^{-}\right){}^{(0,++)}{\cal A}^{{\cal V}_{12\phi}{\cal V}_{34\phi}}_{\Delta_1 \Delta_2 |\Delta_+| \Delta_3 \Delta_4}\left({\bar k}^+_1,{\bar k}^+_2,{\bar k}^+_3,{\bar k}^+_4\right), 
\end{multline}
which, in turn, using the analytic continuations \eqref{BDtoEAdS} can be written in terms of the corresponding exchange diagram in EAdS \cite{Sleight:2020obc,Sleight:2021plv}:
\begin{multline}
{}^{(0,++)}{\cal A}^{{\cal V}_{12\phi}{\cal V}_{34\phi}}_{\Delta_1 \Delta_2 |\Delta_+| \Delta_3 \Delta_4}\left(k_1,k_2,k_3,k_4\right) \\ = \left(\prod^4_{i=1}c^{\text{dS-AdS}}_{\Delta_i}\left(-\eta_0\right)^{\Delta_i}\right)c^{\text{dS-AdS}}_{\Delta_+} e^{-\left(\frac{-d+\Delta_++\Delta_1+\Delta_2}{2}\right)\pi i}e^{-\left(\frac{-d+\Delta_++\Delta_3+\Delta_4}{2}\right)\pi i}\\ \times {\cal A}^{\text{AdS}}_{\Delta_1 \Delta_2 |\Delta_+| \Delta_3 \Delta_4}\left(k_1,k_2,k_3,k_4\right), 
\end{multline}
where 
\begin{multline}
{\cal A}^{\text{AdS}}_{\Delta_1 \Delta_2 |\Delta_+| \Delta_3 \Delta_4}\left(k_1,k_2,k_3,k_4\right)= g_{12} g_{34} \int_0^{\infty} \frac{{\rm d}z_1}{z_1^{d+1}}\int_0^{+\infty} \frac{{\rm d}z_2}{z_2^{d+1}}\\ 
 \times K^{\text{AdS}}_{\Delta_1}\left(z_1,k_1\right)K^{\text{AdS}}_{\Delta_2}\left(z_1,k_2\right)G^{\text{AdS}}_{\Delta_+}(z_1,k;z_2,k)
K^{\text{AdS}}_{\Delta_3}\left(z_2,k_3\right)K^{\text{AdS}}_{\Delta_4}\left(z_2,k_4\right).
\end{multline}

\vskip 4pt
The above procedure can be repeated for all other contributions (i.e. $+-$, $-+$, $--$) to the exchange in the Schwinger-Keldysh formalism, expressing the result in terms of corresponding exchange diagrams in EAdS with appropriate rotations of the external momenta. This expression for the full de Sitter exchange can be found in the \textit{Mathematica} file at \cite{GitHub}.

\section{Conformal Ward Identities}

\label{sec::CWI}

The subgroup $SO^{+}(1,d+1)$ of the de Sitter isometry group acts on the boundary at infinity like the connected conformal group, which implies that late-time correlation functions in dS invariant vacua are subject to conformal Ward identities. In following we discuss how, for a generic $\alpha$-vacuum, the results of the previous sections fit into this picture, both in position and momentum space. The corresponding story in Mellin space \cite{Sleight:2019mgd,Sleight:2019hfp,Sleight:2020obc,Sleight:2021plv} is discussed in appendix \ref{subsec::MSextalpha}.

\vskip 4pt
dS boundary correlators have been extensively studied as solutions to conformal Ward identities in momentum space \cite{Maldacena:2011nz,Mata:2012bx,Bzowski:2013sza,Ghosh:2014kba,Kundu:2014gxa,Kundu:2015xta,Bzowski:2015pba,Bzowski:2015yxv,Shukla:2016bnu,Bzowski:2018fql,Arkani-Hamed:2018kmz,Isono:2019ihz,Isono:2019wex,Bzowski:2019kwd,Baumann:2019oyu,Baumann:2020dch,Bzowski:2022rlz,Bzowski:2023nef}, largely motivated by potential applications in Inflationary Cosmology. It is well known \cite{Bzowski:2013sza,Arkani-Hamed:2018kmz} that boundary correlators in a generic $\alpha$-vacuum with non-zero $\alpha$ require us to admit solutions to the conformal Ward identities that exhibit singularities for collinear momentum configurations. Consider for example three-point functions of scalar operators with dimensions $\Delta_i$. Momentum space conformal Ward identities are reviewed in appendix \ref{subsec::ABP} and reduce to a system \eqref{SKWI} of two second order partial differential equations in the magnitudes $k_i$, $i=1,2,3$. There are therefore four independent solutions and the one with no collinear singularities admits the ``triple-$K$" form \cite{Bzowski:2013sza}:
\begin{equation}\label{3K}
    {}^{(0)}F^\prime_{\Delta_1 \Delta_2 \Delta_3}\left(k_1,k_2,k_3\right)= \int^{\infty}_{0}\frac{{\rm d}z}{z^{d+1}}\,K^{\text{AdS}}_{\Delta_1}\left(z,k_1\right)K^{\text{AdS}}_{\Delta_2}\left(z,k_2\right)K^{\text{AdS}}_{\Delta_3}\left(z,k_3\right),
\end{equation}
where the prime denotes the fact that we stripped off the overall momentum conserving delta function. This expression converges for physical momentum configurations $k_1+k_2+k_3 >0$ and is singular for $k_1+k_2+k_3=0$.\footnote{Interpreted in dS space, this singularity arises from interactions in the short-distance (early-time) limit and the residue is the (UV-limit of the) corresponding flat-space amplitude \cite{Maldacena:2011nz,Raju:2012zr}. Note that this singularity cannot be accessed for physical momentum configurations and can only be reached by analytically continuing the magnitudes $k_i$ to negative values.} The remaining three independent solutions are obtained by sending $k_i \to {\bar k}_i=-k_i$, under which the conformal Ward identities \eqref{Dward} and \eqref{SKWI} are invariant. These are:
\begin{subequations}\label{folded3pt}
\begin{align}
{}^{(1)}F^\prime_{\Delta_1 \Delta_2 \Delta_3}\left(k_1,k_2,k_3\right)&={}^{(0)}F^\prime_{\Delta_1 \Delta_2 \Delta_3}({\bar k}_1,k_2,k_3),\\
{}^{(2)}F^\prime_{\Delta_1 \Delta_2 \Delta_3}\left(k_1,k_2,k_3\right)&={}^{(0)}F^\prime_{\Delta_1 \Delta_2 \Delta_3}(k_1,{\bar k}_2,k_3),\\
{}^{(3)}F^\prime_{\Delta_1 \Delta_2 \Delta_3}\left(k_1,k_2,k_3\right)&={}^{(0)}F^\prime_{\Delta_1 \Delta_2 \Delta_3}(k_1,k_2,{\bar k}_3),
\end{align}
\end{subequations}
which are singular for the respective collinear configurations: ${\bar k}_1+k_2+k_3=0$, $k_1+{\bar k}_2+k_3=0$ and $k_1+k_2+{\bar k}_3=0$. The general solution to the momentum space conformal Ward identities therefore takes the form
\begin{multline}
F^\prime_{\Delta_1 \Delta_2 \Delta_3}\left(k_1,k_2,k_3\right) = a_{\Delta_1 \Delta_2 \Delta_3}{}^{(0)}F^\prime_{\Delta_1 \Delta_2 \Delta_3}\left(k_1,k_2,k_3\right)+b_{\Delta_1 \Delta_2 \Delta_3}{}^{(0)}F^\prime_{\Delta_1 \Delta_2 \Delta_3}({\bar k}_1,k_2,k_3)\\+b_{\Delta_2 \Delta_1 \Delta_3}{}^{(0)}F^\prime_{\Delta_1 \Delta_2 \Delta_3}(k_1,{\bar k}_2,k_3)+b_{\Delta_3 \Delta_2 \Delta_1}{}^{(0)}F^\prime_{\Delta_1 \Delta_2 \Delta_3}(k_1,k_2,{\bar k}_3),
\end{multline}
with undetermined constants $a_{\Delta_1 \Delta_2 \Delta_3}$ and $b_{\Delta_1 \Delta_2 \Delta_3}$, where Bose symmetry requires that $a_{\Delta_1 \Delta_2 \Delta_3}$ is symmetric in $\Delta_1$, $\Delta_2$ and $\Delta_3$. Indeed, the three-point contact diagram \eqref{3ptcontactdS} in a generic $\alpha$-vacuum takes this form, where the time-ordering of the contribution dictates whether ${\bar k} = {\bar k}^-$ or ${\bar k}^+$. The contact diagrams in the Bunch-Davies vacuum \eqref{BD3pt} correspond to $b_{\Delta_1 \Delta_2 \Delta_3}=0$, as anticipated.

\vskip 4pt
In contrast, there is scarce literature on how boundary correlators in a generic dS invariant vacuum might arise as solutions to position space conformal Ward identities. In the Bunch-Davies vacuum, three-point functions in position space take the following form in the embedding formalism:
\begin{align}\nonumber
    {}^{(0)}F^\prime_{\Delta_1 \Delta_2 \Delta_3}\left(Q_1,Q_2,Q_3\right)&= \int {\rm d}^{d+1}X_{\text{AdS}^+}\,K^{\text{AdS}}_{\Delta_1}(X_{\text{AdS}^+},Q_1)K^{\text{AdS}}_{\Delta_2}(X_{\text{AdS}^+},Q_2) K^{\text{AdS}}_{\Delta_3}(X_{\text{AdS}^+},Q_3),\\
    & = \frac{C_{\Delta_1\Delta_2\Delta_3} }{\left(-2Q_{1}\cdot Q_2\right)^{\frac{\Delta_1+\Delta_2-\Delta_3}{2}}\left(-2Q_{2}\cdot Q_3\right)^{\frac{\Delta_2+\Delta_3-\Delta_1}{2}}\left(-2Q_{3}\cdot Q_1\right)^{\frac{\Delta_1+\Delta_3-\Delta_1}{2}}}, \label{BDpos}
\end{align}
for some symmetric coefficient $C_{\Delta_1\Delta_2\Delta_3}$. This is the standard form required by symmetry under the connected component of the conformal group for correlators with separated points \cite{Polyakov:1970xd,Weinberg:2010fx,Costa:2011mg} and is the Fourier transform of the solution \eqref{3K} as shown in \cite{Bzowski:2013sza}.\footnote{Note that symmetry with respect to global conformal transformations assumes symmetry under inversions and hence under parity transformations (see section \ref{sec::dSST}). To find parity-breaking solutions one should require symmetry with respect to the connected component only. This point however does not affect the discussion for scalars, which are the focus of this work.} To understand how the solutions corresponding to \eqref{folded3pt} could arise in position space, recall that (see section \ref{sec::dSST}) that the antipodal transformation \eqref{ap} is equivalent to the identity from the perspective of the global conformal group and may therefore be used to generate other solutions to the position space conformal Ward identities. Indeed, as shown in section \ref{subsec::alphafromBD}, the rotations of the momenta that appear in the solutions \eqref{folded3pt} are generated by the antipodal transformation of the bulk point in the corresponding propagator. This suggests the following position space interpretation:
\begin{align}\nonumber
    {}^{(1)}F^\prime_{\Delta_1 \Delta_2 \Delta_3}\left(Q_1,Q_2,Q_3\right)&= \int {\rm d}^{d+1}X_{\text{AdS}^+}\,K^{\text{AdS}}_{\Delta_1}(X_{\text{AdS}^+}\cdot {\bar Q}^-_1-i\epsilon)K^{\text{AdS}}_{\Delta_2}(X_{\text{AdS}^+},Q_2) K^{\text{AdS}}_{\Delta_3}(X_{\text{AdS}^+},Q_3),\\
    & = \frac{C_{\Delta_1\Delta_2\Delta_3} }{(-2{\bar Q}^-_1\cdot Q_2+i\epsilon)^{\frac{\Delta_1+\Delta_2-\Delta_3}{2}}(-2Q_{2}\cdot Q_3)^{\frac{\Delta_2+\Delta_3-\Delta_1}{2}}(-2Q_{3}\cdot {\bar Q}^-_1+i\epsilon)^{\frac{\Delta_1+\Delta_3-\Delta_1}{2}}},
\end{align}
where we replaced the antipodal transformation of $X_{\text{AdS}^+}$ in the bulk-to-boundary propagator $K^{\text{AdS}}_{\Delta_1}$ with one of the boundary point $Q_1$. Likewise:
\begin{align}
{}^{(2)}F^\prime_{\Delta_1 \Delta_2 \Delta_3}\left(Q_1,Q_2,Q_3\right)&=\frac{C_{\Delta_1\Delta_2\Delta_3} }{(-2Q_1\cdot {\bar Q}^-_2+i\epsilon)^{\frac{\Delta_1+\Delta_2-\Delta_3}{2}}(-2{\bar Q}^-_{2}\cdot Q_3+i\epsilon)^{\frac{\Delta_2+\Delta_3-\Delta_1}{2}}(-2Q_{3}\cdot Q_1)^{\frac{\Delta_1+\Delta_3-\Delta_1}{2}}},\\
{}^{(3)}F^\prime_{\Delta_1 \Delta_2 \Delta_3}\left(Q_1,Q_2,Q_3\right)&=\frac{C_{\Delta_1\Delta_2\Delta_3} }{(-2Q_1\cdot Q_2)^{\frac{\Delta_1+\Delta_2-\Delta_3}{2}}(-2Q_{2}\cdot {\bar Q}^-_3+i\epsilon)^{\frac{\Delta_2+\Delta_3-\Delta_1}{2}}(-2{\bar Q}^-_{3}\cdot Q_1+i\epsilon)^{\frac{\Delta_1+\Delta_3-\Delta_1}{2}}}.
\end{align}

\vskip 4pt
The above discussion naturally extends to higher-point conformal correlation functions, where other solutions to the conformal Ward identities can be generated by application of the antipodal transformation \eqref{ap} to the boundary points $Q_i$, or equivalently, flipping the sign of the magnitudes $k_i$ in momentum space. This is supported by the structure of the perturbative bulk results for generic dS invariant vacua in the previous sections.

\vskip 4pt
Let us conclude this section by noting that such additional solutions to the conformal Ward identities are discarded in the context of holographic correlators in the AdS/CFT correspondence, where in Euclidean signature one considers bulk points that lie on the upper sheet of the EAdS hyperboloid (see figure \ref{fig::ds2ads}).  Note that in the Lorentzian picture this issue does not arise at all since the boundary of the AdS hyperboloid is connected. For this reason, the additional solutions in this section do not play a role in the context of Lorentzian CFTs and are an intrinsically Euclidean feature.

\section{Summary and Outlook}

In this work we initiated the study of late-time correlators in de Sitter space for Bogoliubov initial states through their reformulation in terms of their Bunch-Davies counterparts, and in turn, Witten diagrams in Euclidean anti-de Sitter space. Possible directions for future work include:

\begin{itemize}
\item {\bf Fields with spin.}  While we focused on scalar field theories, the extension to fields of non-zero spin should be straightforward. Indeed, the perturbative reformulation of late-time correlators in the Bunch-Davies vacuum in terms of their EAdS counterparts has also been derived for fields of arbitrary integer spin \cite{Sleight:2019hfp,Sleight:2020obc,Sleight:2021plv} and Fermions \cite{Schaub:2023scu}. 

\item {\bf Inflationary correlators.} In slow roll inflation it is possible to obtain inflationary correlators by perturbing late-time correlators of exact de Sitter space \cite{Creminelli:2003iq,Arkani-Hamed:2015bza,Arkani-Hamed:2018kmz}. It would be interesting to revisit the apparent violation \cite{Shukla:2016bnu,Ansari:2024pgq} of the Maldacena consistency condition for three-point function of scalar perturbations in $\alpha$-vacua using the approach presented in this work, especially in the view of the recent positive results \cite{Ansari:2024pgq} at four-point and the structure of the OPE limit for $\alpha$-vacua discussed in this work. 

\item {\bf More general initial states.} It would be interesting to use the techniques presented in this work to examine the properties of late-time correlators in Bogoliubov initial states in more detail and to explore whether they might be extended to more general states that are not the Bogoliubov transform of the Bunch-Davies vacuum.

\item {\bf Celestial correlators.} Holographic correlation functions on the celestial sphere of Minkowski space have recently been defined by considering the Mellin transform of time-ordered Minkowski correlation functions with respect to the radial direction in the hyperbolic slicing \cite{Sleight:2023ojm,Iacobacci:2024nhw}. The effective reduced vacuum on the de Sitter slicing is the Bunch-Davies vacuum and accordingly such celestial correlation functions can be re-written in terms of Witten diagrams in EAdS, recycling the results \cite{Sleight:2020obc,Sleight:2021plv} for the Bunch-Davies vacuum. If one instead considers Lorentz (not Poincar\'e) invariant vacua, the effective reduced vacua on the de Sitter slicing are $\alpha$-vacua \cite{Melton:2023dee}. It would therefore be interesting to understand whether celestial correlators in Lorentz (not Poincar\'e) invariant vacua can be studied along similar lines using the results of this work.
\end{itemize}

\section*{Acknowledgments}

We thank Ashish Shukla for collaboration at an early stage of this work. We thank Paolo Benincasa, Guilherme L. Pimentel, Ashish Shukla, Kostas Skenderis and Sandip Trivedi for discussions. The research of CS and MT was supported by the European Union (ERC grant ``HoloBoot'', project number 101125112),\footnote{Views and opinions expressed are however those of the author(s) only and do not necessarily reflect those of the European Union or the European Research Council. Neither the European Union nor the granting authority can be held responsible for them.} by the MUR-PRIN grant No. PRIN2022BP52A (European Union - Next Generation EU) and by the INFN initiative STEFI. The research of AJC and CS was partially supported by the STFC grant ST/T000708/1.

\newpage

\begin{appendix}

\section{Mellin Space}
\label{app::Mellin}

In the presence of a translation symmetry it is often convenient to work in Fourier space i.e. in a basis of plane waves $e^{\pm ikx}$ that diagonalise the translation generator and thus make the symmetry manifest: For functions $f\left(x\right)$ in $L^2\left(\mathbb{R},\text{d}x\right)$ we expand
\begin{equation}
    f\left(x\right)=\int^\infty_{-\infty}\frac{{\rm d}k}{2\pi}f(k)\,e^{ikx},
\end{equation}
where
\begin{equation}
f\left(k\right)=\int^\infty_{-\infty}{\rm d}x f(x)\,e^{-ikx}.
\end{equation}
For this reason, in de Sitter space one often considers late-time correlators in Fourier-space due to the translation symmetry in the spatial (boundary) directions. The remaining direction is the bulk time direction, where there is no translation symmetry owing to the time dependence of the de Sitter background and there is therefore less benefit from transforming to Fourier space. 

\vskip 4pt
In de Sitter space there is also symmetry under dilatations $(\eta,\vec{x}) \to (\lambda \eta, \lambda \vec{x})$, for which it is instead natural to work in Mellin space \cite{Sleight:2019mgd,Sleight:2019hfp,Sleight:2020obc,Sleight:2021plv}. For functions $f(\eta)$ in $L^2\left(\mathbb{R}^+,\frac{{\rm d}(-\eta)}{(-\eta)^{d+1}}\right)$ we expand
\begin{subequations}
 \begin{align}
    f(\eta)&=\int^{i\infty}_{-i\infty}\frac{{\rm d}s}{2\pi i}2 f(s)\left(-\eta\right)^{-\left(2s-\tfrac{d}{2}\right)},\\
    f(s)&=\int^0_{-\infty}\frac{{\rm d}\eta}{-\eta} f(\eta)\left(-\eta\right)^{2s-\tfrac{d}{2}},
\end{align}   
\end{subequations}
where the monomials $\left(-\eta\right)^{\mp\left(2s-\tfrac{d}{2}\right)}$ are analogous to the plane waves in Fourier space and in this case diagonalise the dilatation generator. The same holds for Euclidean AdS with respect to the bulk Poincar\'e coordinate $z$. For this reason bulk integrals encountered in the perturbative computation of boundary correlators are trivialised in Mellin space, where they take the form
\begin{equation}\label{DDMellin}
\int^{0}_{-\infty}\frac{{\rm d}\left(-\eta\right)}{\left(-\eta\right)^{d+1}}\left(-\eta\right)^{D} \left(-\eta\right)^{-\sum\limits_{i}\left(2s_i-\tfrac{d}{2}\right)} = 2\pi i\, \delta\left(D-d-\sum\limits_{i}\left(2s_i-\tfrac{d}{2}\right)\right),
\end{equation}
where $D$ parameterises any powers of $\eta$ generated by a derivative interaction.  This is analogous to momentum conservation in the presence of translation symmetry and, as reviewed in section \ref{Dward}, is a consequence of the Dilatation Ward identity.

\vskip 4pt
In Mellin space the analytic structure of a function in $k$ and $\eta$ is encoded by the pole structure of the Mellin transform and for this reason makes manifest certain identities satisfied by propagators and their corresponding boundary correlators, both in dS and AdS - as well as the relationship between the two under analytic continuation. This will be detailed in sections \ref{subsec::EAdSrotMB}, \ref{app::PI} and \ref{app::CI} for the examples encountered in this work. In section \ref{subsec::MSextalpha} we discuss the structure of boundary correlators in a generic $\alpha$-vacuum in Mellin space. In section \ref{app::MBreg} we will discuss singularities of Mellin integrals generated by pole pinching and their regularisation.

\subsection{Rotation to EAdS}
\label{subsec::EAdSrotMB}

As noted in \cite{Sleight:2019mgd,Sleight:2019hfp,Sleight:2020obc,Sleight:2021plv}, Mellin space makes manifest that dS boundary correlators can be perturbatively re-cast in terms of corresponding boundary correlators in EAdS. This is reviewed below for the Bunch-Davies vacuum $\left(\alpha =0\right)$, which we extend to generic $\alpha$ in section \ref{subsec::MSextalpha}.

\vskip 4pt
In Mellin space, the bulk-to-bulk propagators for boundary correlators in dS and EAdS can be expressed in the form \cite{Sleight:2020obc,Sleight:2021plv}:\footnote{In \eqref{bubuMB} we introduced the notation 
  \begin{equation}
      [{\rm d}u]_n:=\frac{{\rm d}u_1}{2\pi i}\ldots\frac{{\rm d}u_n}{2\pi i}.
  \end{equation}}
\begin{align}\nonumber
  \hspace*{-0.5cm}  G^{(0)}_{\pm {\hat \pm}}(\eta_1,k_1;\eta_2,k_2)&=\int^{i\infty}_{-i\infty}[{\rm d}u]_2\, G^{(0)}_{\pm {\hat \pm}}(u_1,k_1;u_2,k_2)\left(-\eta_1\right)^{-\left(2u_1-\tfrac{d}{2}\right)}\left(-\eta_2\right)^{-\left(2u_2-\tfrac{d}{2}\right)},\\ \label{bubuMB}
  G^{\text{AdS}}_{\Delta}(z_1,k_1;z_2,k_2)&=\int^{i\infty}_{-i\infty}[{\rm d}u]_2\, G^{\text{AdS}}_{\Delta}(u_1,k_1;u_2,k_2)z_1^{-\left(2u_1-\tfrac{d}{2}\right)}z_2^{-\left(2u_2-\tfrac{d}{2}\right)},  
   \end{align}    
where
  \begin{subequations}\label{bubuMBrep}
 \begin{align}\nonumber
      G^{(0)}_{\pm {\hat \pm}}(u_1,k_1;u_2,k_2)&= \frac{2\pi i}{\nu} \csc\left(\pi(u_1+u_2)\right)\left(e^{{\hat \pm} \pi \nu}
  \omega_{\Delta_+}(u_1,u_2)-e^{\mp \pi \nu}\omega_{\Delta_-}(u_1,u_2)\right)\\ & \hspace*{6cm} \times \Omega^{\pm\,{\hat \pm}}_{\nu}(u_1,k_1;u_2,k_2),\label{bubuMBrepdS}\\
      G^{\text{AdS}}_{\Delta}(u_1,k_1;u_2,k_2)&= \csc\left(\pi(u_1+u_2)\right)
  \omega_{\Delta}(u_1,u_2) \Gamma(+i\nu)\Gamma(-i\nu) \nonumber \\ & \hspace*{6cm} \times\Omega^{\text{AdS}}_{\nu}(u_1,k_1;u_2,k_2)\label{bubuMBrepAdS},
  \end{align}     
  \end{subequations}
and
   \begin{subequations}\label{app::splitrepdS}
  \begin{align}
        \Omega^{\pm\,{\hat \pm}}_{\nu}(u_1,k_1;u_2,k_2)&=\frac{\nu^2}{\pi}K^{(0)}_{\Delta_+,\pm}\left(u_1,k_1\right)K^{(0)}_{\Delta_-,{\hat \pm}}\left(u_2,k_2\right),\\
        \Omega^{\text{AdS}}_{\nu}(u_1,k_1;u_2,k_2)&=\frac{\nu^2}{\pi}K^{\text{AdS}}_{\Delta_+}\left(u_1,k_1\right)K^{\text{AdS}}_{\Delta_-}\left(u_2,k_2\right).
\end{align}
  \end{subequations}
  The latter are a product of the corresponding bulk-to-boundary propagators, which in Mellin space read
  \begin{subequations}\label{buboMB}
   \begin{align}
K^{(0)}_{\Delta,\pm}\left(\eta,k\right)&=\int^{i\infty}_{-i\infty}\frac{{\rm d}s}{2\pi i}K^{(0)}_{\Delta,\pm}\left(s,k\right)\left(-\eta\right)^{-\left(2s-\tfrac{d}{2}\right)},\\
K^{\text{AdS}}_{\Delta}\left(z,k\right)&=\int^{i\infty}_{-i\infty}\frac{{\rm d}s}{2\pi i}K^{\text{AdS}}_{\Delta}\left(s,k\right)z^{-\left(2s-\tfrac{d}{2}\right)},
\end{align}       
  \end{subequations}
where 
\begin{subequations}
 \begin{align}
K^{(0)}_{\Delta,\pm}\left(s,k\right)&=\left(-\eta_0\right)^{\Delta}\frac{\Gamma(-i\nu)}{4\pi}e^{\mp \left(s+\tfrac{i\nu}{2}\right)\pi i}\Gamma(s+\tfrac{i\nu}{2})\Gamma(s-\tfrac{i\nu}{2})\left(\frac{k}{2}\right)^{-2s+i\nu},\label{KdSmellin}\\
K^{\text{AdS}}_{\Delta}\left(s,k\right)&=\frac{1}{2\Gamma\left(i\nu+1\right)}\Gamma(s+\tfrac{i\nu}{2})\Gamma(s-\tfrac{i\nu}{2})\left(\frac{k}{2}\right)^{-2s+i\nu}.\label{KAdSmellin}
\end{align}   
\end{subequations}
 The functions
\begin{equation}\label{projectors}
    \omega_{\Delta_\pm}(u_1,u_2)=2\sin(\pi(u_1 \mp \tfrac{i\nu}{2}))\sin(\pi(u_2 \mp \tfrac{i\nu}{2})),
\end{equation}
serve to project onto the $\Delta_\pm$ contributions and the factor $\csc\left(\pi(u_1+u_2)\right)$ encodes the contact terms. A more detailed explanation of this representation for (EA)dS propagators and how it is obtained can be found in \cite{Sleight:2020obc,Sleight:2021plv}.

\vskip 4pt
The relationship between boundary correlators in dS and EAdS emerges upon noting that:
\begin{subequations}
 \begin{align}
  \hspace*{-0.5cm}  \Omega^{\pm\,{\hat \pm}}_{\nu}(u_1,k_1;u_2,k_2)&=c^{\text{dS-AdS}}_{\frac{d}{2}+i\nu}c^{\text{dS-AdS}}_{\frac{d}{2}-i\nu}e^{\mp \left(u_1+\tfrac{i\nu}{2}\right)\pi i}e^{{\hat \mp} \left(u_2-\tfrac{i\nu}{2}\right)\pi i}\Omega^{\text{AdS}}_{\nu}(u_1,k_1;u_2,k_2),\\
   K^{(0)}_{\Delta,\pm}\left(s,k\right)&=c^{\text{dS-AdS}}_{\Delta} e^{\mp \left(s+\tfrac{i\nu}{2}\right)\pi i}K^{\text{AdS}}_{\Delta}\left(s,k\right),
\end{align}   
\end{subequations}
where the relative phases in Mellin space make manifest the analytic continuations \eqref{BDtoEAdS} to EAdS:
\begin{subequations}
    \begin{align}
         G^{(0)}_{\pm {\hat \pm}}(\eta_1,\eta_2)&\to c^{\text{dS-AdS}}_{\Delta_+}e^{\mp \frac{\Delta_+\pi i}{2}}e^{{\hat \mp} \frac{\Delta_+\pi i}{2}}G^{\text{AdS}}_{\Delta_+}(z_1,z_2)+\,\left(\Delta_+ \to \Delta_-\right),\\
 K^{(0)}_{\Delta,\,
 \pm}(\eta,k)&\to c^{\text{dS-AdS}}_{\Delta}e^{\mp \frac{\Delta \pi i}{2}}K^{\text{AdS}}_{\Delta}(z,k),
    \end{align}
\end{subequations}
under
\begin{subequations}
 \begin{align}
    \emph{time-ordered points}: \quad &\eta\, \to\, +iz,\\
    \emph{anti-time-ordered points}: \quad &\eta\, \to\, -iz.
\end{align}   
\end{subequations}

\subsection{Propagator identities}
\label{app::PI}

In this section we use Mellin space to prove various propagator identities appearing in the main text.

\paragraph{Bulk-to-bulk propagators.} For the identities \eqref{TaTBD} expressing (anti-)time-ordered propagators in a generic $\alpha$ vacuum in terms of their Bunch-Davies counterparts, the starting point is the definition:
\begin{subequations}\label{appdefGTbT}
  \begin{align}
    G^{(\alpha)}_{T}\left(\eta_1;\eta_2\right)&=\theta\left(\eta_1-\eta_2\right)G^{(\alpha)}_{W}\left(\eta_1;\eta_2\right)+\theta\left(\eta_2-\eta_1\right)G^{(\alpha)}_{W}\left(\eta_2;\eta_1\right),\\
    G^{(\alpha)}_{{\bar T}}\left(\eta_1;\eta_2\right)&=\theta\left(\eta_1-\eta_2\right)G^{(\alpha)}_{W}\left(\eta_2;\eta_1\right)+\theta\left(\eta_2-\eta_1\right)G^{(\alpha)}_{W}\left(\eta_1;\eta_2\right),
\end{align}  
\end{subequations}
which we combine with the expression \eqref{wafrombd} for the Wightman function $G^{(\alpha)}_{W}$ in terms of its Bunch-Davies counterpart $G^{(0)}_{W}$, which we repeat below for convenience:
\begin{multline}\label{app::wafrombd}
G^{(\alpha)}_{W}(\eta_1;\eta_2) = \cosh^2\alpha \, G^{(0)}_{W}(\eta_1;\eta_2)+\sinh^2\alpha \, G^{(0)}_{W}({\bar \eta}^{-}_1;{\bar \eta}^+_2)\\-\frac{1}{2}\sinh 2\alpha [ e^{i\left(\beta+\frac{\pi d}{2}\right)} \, G^{(0)}_{W}(\eta_1;{\bar \eta}^+_2)+ e^{-i\left(\beta+\frac{\pi d}{2}\right)} \, G^{(0)}_{W}({\bar \eta}^-_1;\eta_2)]. 
\end{multline}
Using the Mellin-Barnes representation \eqref{bubuMB} it is straightforward to show that: 
\begin{align} \nonumber 
   G^{(0)}_{W}({\bar \eta}^{-}_1;{\bar \eta}^+_2)&=\int^{i\infty}_{-i\infty}[{\rm d}u]_2\, G^{(0)}_{-+}(u_1,k_1;u_2,k_2)\left(-\eta_1 e^{+\pi i}\right)^{-\left(2u_1-\tfrac{d}{2}\right)}\left(-\eta_2 e^{-\pi i}\right)^{-\left(2u_2-\tfrac{d}{2}\right)},\\ \nonumber
   &=\int^{i\infty}_{-i\infty}[{\rm d}u]_2\, G^{(0)}_{+-}(u_1,k_1;u_2,k_2)\left(-\eta_1\right)^{-\left(2u_1-\tfrac{d}{2}\right)}\left(-\eta_2 \right)^{-\left(2u_2-\tfrac{d}{2}\right)},\\
   &=G^{(0)}_{W}(\eta_2;\eta_1).
\end{align}
In the same way, one also obtains:
\begin{subequations}
 \begin{align}
    G^{(0)}_{\Delta,T}\left(\eta_1;\eta_2\right)&=e^{-2\Delta \pi i}G^{(0)}_{\Delta,{\bar T}}\left({\bar \eta}^-_1;{\bar \eta}^-_2\right),\\
    G^{(0)}_{\Delta,{\bar T}}\left(\eta_1;\eta_2\right)&=e^{+2\Delta \pi i}G^{(0)}_{\Delta,T}\left({\bar \eta}^+_1;{\bar \eta}^+_2\right),
\end{align}   
\end{subequations}
and
\begin{subequations}
 \begin{align}
    G^{(0)}_{W}\left(\eta_1;{\bar \eta}^+_2\right)&=e^{+\Delta_+ \pi i}G^{(0)}_{\Delta_+,T}\left({\bar \eta}^+_1;{\bar \eta}^+_2\right)+e^{+\Delta_- \pi i}G^{(0)}_{\Delta_-,T}\left({\bar \eta}^+_1;{\bar \eta}^+_2\right),\\
    &=e^{-\Delta_+ \pi i}G^{(0)}_{\Delta_+,{\bar T}}\left(\eta_1;\eta_2\right)+e^{-\Delta_- \pi i}G^{(0)}_{\Delta_-,{\bar T}}\left(\eta_1;\eta_2\right),
\end{align}   
\end{subequations}
\begin{subequations}
 \begin{align}
    G^{(0)}_{W}\left({\bar \eta}^-_1;\eta_2\right)&=e^{-\Delta_+ \pi i}G^{(0)}_{\Delta_+,T}\left({\bar \eta}^-_1;{\bar \eta}^-_2\right)+e^{-\Delta_- \pi i}G^{(0)}_{\Delta_-,T}\left({\bar \eta}^-_1;{\bar \eta}^-_2\right),\\
    &=e^{\Delta_+ \pi i}G^{(0)}_{\Delta_+,{\bar T}}\left(\eta_1;\eta_2\right)+e^{\Delta_- \pi i}G^{(0)}_{\Delta_-,{\bar T}}\left(\eta_1;\eta_2\right).
\end{align}   
\end{subequations}

Plugging everything into the definition \eqref{appdefGTbT} one obtains:
\begin{align}\nonumber
    G^{(\alpha)}_{T}\left(\eta_1;\eta_2\right)&=\cosh^2\alpha\,G^{(0)}_{T}\left(\eta_1;\eta_2\right)+\sinh^2\alpha\,G^{(0)}_{{\bar T}}\left(\eta_1;\eta_2\right) \\
    &\hspace*{2.5cm}-\frac{1}{2}\sinh 2\alpha [ e^{i\left(\beta+\frac{\pi d}{2}\right)} \, G^{(0)}_{W}(\eta_1;{\bar \eta}^+_2)+ e^{-i\left(\beta+\frac{\pi d}{2}\right)} \, G^{(0)}_{W}({\bar \eta}^-_1;\eta_2)], \nonumber\\
    &=P^+_{\Delta} G^{(0)}_{\Delta_+,T}(\eta_1;\eta_2)+e^{2\Delta_+ \pi i}P^-_{\Delta} G^{(0)}_{\Delta_+,T}({\bar \eta}^+_1;{\bar \eta}^+_2) + (\Delta_+ \to \Delta_-),
\end{align}
and 
\begin{align}\nonumber
    G^{(\alpha)}_{{\bar T}}\left(\eta_1;\eta_2\right)&=\cosh^2\alpha\,G^{(0)}_{{\bar T}}\left(\eta_1;\eta_2\right)+\sinh^2\alpha\,G^{(0)}_{T}\left(\eta_1;\eta_2\right) \\
    &\hspace*{2.5cm}-\frac{1}{2}\sinh 2\alpha [ e^{i\left(\beta+\frac{\pi d}{2}\right)} \, G^{(0)}_{W}(\eta_1;{\bar \eta}^+_2)+ e^{-i\left(\beta+\frac{\pi d}{2}\right)} \, G^{(0)}_{W}({\bar \eta}^-_1;\eta_2)], \nonumber\\
    &=M^-_{\Delta} G^{(0)}_{\Delta_+,{\bar T}}(\eta_1;\eta_2)+ \, e^{-2\Delta_+ \pi i} M^+_{\Delta} G^{(0)}_{\Delta_+,{\bar T}}({\bar \eta}^-_1;{\bar \eta}^-_2)+ (\Delta_+ \to \Delta_-). 
\end{align}

\vskip 4pt

\paragraph{Bulk-to-boundary propagators.} Bulk-to-boundary propagators are obtained by taking the late-time limit $\eta \to 0$ of one of the bulk points in the free theory two-point functions, which is simple to perform in the Mellin-Barnes representation \eqref{bubuMB}. The expansion for small $\eta_2$ is generated by the residues of the poles in the corresponding Mellin variable $u_2$:
\begin{equation}
    u_2 = \pm \tfrac{i\nu}{2}-n, \qquad n=0,\,1,\,2,\,\ldots\,,
\end{equation}
where the leading terms are encoded the $n=0$ poles. In particular:
\begin{align}
   \lim_{\eta_2\to 0} G^{(\alpha)}_{\pm \mp}(\eta_1;\eta_2) = K^{(\alpha)}_{\Delta_+,\pm}(\eta_1,k)+K^{(\alpha)}_{\Delta_-,\pm}(\eta_1,k),
\end{align}
which identifies the bulk-to-boundary propagators:
\begin{subequations}\label{appbuboasres}
 \begin{multline}
    K^{(\alpha)}_{\Delta_\pm,+}(\eta_1,k)=\int^{+i\infty}_{-i\infty}\frac{{\rm d}u_1}{2\pi i}\,\text{Res}_{u_2=\mp \tfrac{i\nu}{2}}\left[G^{(\alpha)}_{+ -}(u_1,u_2)\left(-\eta_2\right)^{-\left(2u_2-\tfrac{d}{2}\right)}\right]\\ \times \left(-\eta_1\right)^{-\left(2u_1-\tfrac{d}{2}\right)},
   \end{multline}   
 \begin{multline}
    K^{(\alpha)}_{\Delta_\pm,-}(\eta_1,k)=\int^{+i\infty}_{-i\infty}\frac{{\rm d}u_1}{2\pi i}\,\text{Res}_{u_2=\mp \tfrac{i\nu}{2}}\left[G^{(\alpha)}_{-+}(u_1,u_2)\left(-\eta_2\right)^{-\left(2u_2-\tfrac{d}{2}\right)}\right]\\ \times \left(-\eta_1\right)^{-\left(2u_1-\tfrac{d}{2}\right)}.
\end{multline}   
\end{subequations}
For $\alpha=0$ these recover the Bunch-Davies result \eqref{buboMB}. For generic $\alpha$, combining the identity \eqref{app::wafrombd} with \eqref{appbuboasres} above one obtains \eqref{buboafrombd}:
\begin{subequations}
 \begin{align}
K^{(\alpha)}_{\Delta,\, +}\left(\eta,k\right) &= P^+_{\Delta} K^{(0)}_{\Delta,\, +}(\eta,k) + P^-_{\Delta} e^{\Delta \pi i}K^{(0)}_{\Delta,\, +}({\bar \eta}^+,k),\\
K^{(\alpha)}_{\Delta,\, -}(\eta,k) &= M^+_{\Delta} e^{-\Delta \pi i}K^{(0)}_{\Delta,\, -}({\bar \eta}^-,k) + M^-_{\Delta} K^{(0)}_{\Delta,\, -}(\eta,k).
\end{align}   
\end{subequations}

\vskip 4pt

\paragraph{Boundary two-point function.}

In the same way one obtains the free theory boundary two-point function by taking the late-time limit of both bulk points. Focusing first on the Bunch-Davies case, the non-local contributions (i.e. contributions that are non-analytic in $k$) are given by the following residues:
\begin{align}
   & \lim_{\eta_1,\eta_2 \to 0}G^{(0)}_{\Delta_\pm,W}(\eta_1,k;\eta_2,k)\\ & \hspace*{2cm}=\text{Res}_{u_1=\mp \tfrac{i\nu}{2},\,u_2=\mp \tfrac{i\nu}{2}}\left[G^{(0)}_{\Delta_+,W}(u_1;u_2)\left(-\eta_2\right)^{-\left(2u_2-\tfrac{d}{2}\right)}\left(-\eta_2\right)^{-\left(2u_2-\tfrac{d}{2}\right)}\right] + \ldots\,, \nonumber
    \\ & \hspace*{2cm} =\frac{\left(\eta_1 \eta_2\right)^{\frac{d}{2}}}{4\pi}\Gamma\left(\frac{d}{2}-\Delta_\pm\right)^2\left(\frac{\eta_1\eta_2 k^2}{4}\right)^{\Delta_\pm-\frac{d}{2}}+ \ldots\,, \nonumber
\end{align}
where the $\ldots$ denote local contributions. The result for generic $\alpha$-vacuum then follows from the identity \eqref{wafrombd} expressing the boundary two-point function in a generic $\alpha$ vacuum in terms of the Bunch-Davies result.

\vskip 4pt

\paragraph{Antipodal transformation.} In Mellin space it is straightforward to see that an antipodal transformation corresponds to a rotation of the boundary momentum by a phase $e^{\pm \pi i}$ i.e. equations \eqref{bubudSap} and \eqref{bubodSap}. For bulk-to-boundary propagators \eqref{bubodSap}, in just a few lines we have that:
\begin{subequations}
\begin{align}
K^{(0)}_{\Delta,\pm}({\bar \eta}^{{\hat \pm}},k)&=\int^{i\infty}_{-i\infty}\frac{{\rm d}s}{2\pi i}K^{(0)}_{\Delta,\pm}\left(s,k\right)(-\eta e^{{\hat \mp}\pi i})^{-\left(2s-\tfrac{d}{2}\right)},\\
&=\left(e^{{\hat \pm}\pi i}\right)^{\Delta-d}\int^{i\infty}_{-i\infty}\frac{{\rm d}s}{2\pi i}K^{(0)}_{\Delta,\pm}(s,{\bar k}^{{\hat \pm}})(-\eta)^{-\left(2s-\tfrac{d}{2}\right)},\\
&=\left(e^{{\hat \pm}\pi i}\right)^{\Delta-d}K^{(0)}_{\Delta,\pm}(\eta,{\bar k}^{{\hat \pm}}).
\end{align}    
\end{subequations}
The analogous result \eqref{bubudSap} for the bulk-to-bulk propagators then follows from the representation \eqref{bubuMBrep} in terms of a product of bulk-to-boundary propagators.

\subsection{Boundary correlators and their identities} 
\label{app::CI}

In Mellin space the bulk integrals encountered in the perturbative computation of boundary correlators in (anti-)de Sitter space are trivialised by virtue of the dilatation symmetry. For example, the $n$-point contact diagram generated by the vertex \eqref{npointvertex} in Euclidean AdS$_{d+1}$ in Mellin space is given by
\begin{multline}
{\cal A}^{\text{AdS}}_{\Delta_1 \Delta_2 \ldots \Delta_n}\left(k_1,k_2, \ldots, k_n\right)= \int^{+i\infty}_{-i\infty}[{\rm d}s]_n\,{\cal A}^{\text{AdS}}_{\Delta_1 \Delta_2 \ldots \Delta_n}\left(s_1,k_1;s_2,k_2; \ldots ; s_n,k_n\right),
\end{multline}
where
\begin{align}\label{MBcontact}
{\cal A}^{\text{AdS}}_{\Delta_1 \Delta_2 \ldots \Delta_n}\left(s_1,k_1;s_2,k_2; \ldots ; s_n,k_n\right)&=\int^\infty_0 \frac{{\rm d}z}{z^{d+1}}\,z^{-\sum\limits^n_{i=1}\left(2s_i-\tfrac{d}{2}\right)}\\ & \hspace*{-1.5cm} \times \prod^n_{i=1}\,\frac{\Gamma\left(s_i+\tfrac{1}{2}\left(\tfrac{d}{2}-\Delta_i\right)\right)\Gamma\left(s_i-\tfrac{1}{2}\left(\tfrac{d}{2}-\Delta_i\right)\right)}{2\Gamma\left(\Delta_i-\tfrac{d}{2}+1\right)} \left(\frac{k_i}{2}\right)^{-2s+\Delta-\tfrac{d}{2}}. \nonumber
\end{align}
The integral over the bulk coordinate $z$ generates a Dirac delta function:
\begin{equation}\label{MBdiracdelta}
    \int^\infty_0 \frac{{\rm d}z}{z^{d+1}}\,z^{-\sum\limits^n_{i=1}\left(2s_i-\tfrac{d}{2}\right)} = 2\pi i\, \delta\left(d+\sum\limits^n_{i=1}\left(2s_i-\tfrac{d}{2}\right)\right),
\end{equation}
which constrains the sum of the Mellin variables $s_i$, analogous to momentum conservation in the presence of translation symmetry. The full expression reads
\begin{align}\label{MBcontact}
{\cal A}^{\text{AdS}}_{\Delta_1 \Delta_2 \ldots \Delta_n}\left(s_1,k_1;s_2,k_2; \ldots ; s_n,k_n\right)&=2\pi i\, \delta\left(d+\sum\limits^n_{i=1}\left(2s_i-\tfrac{d}{2}\right)\right)\\ & \hspace*{-1.5cm} \times \prod^n_{i=1}\,\frac{\Gamma\left(s_i+\tfrac{1}{2}\left(\tfrac{d}{2}-\Delta_i\right)\right)\Gamma\left(s_i-\tfrac{1}{2}\left(\tfrac{d}{2}-\Delta_i\right)\right)}{2\Gamma\left(\Delta_i-\tfrac{d}{2}+1\right)} \left(\frac{k_i}{2}\right)^{-2s+\Delta-\tfrac{d}{2}}. \nonumber 
\end{align}

\vskip 4pt
Using the representation \eqref{bubuMBrep} of bulk-to-bulk propagators, the $s$-channel exchange diagram generated by the cubic vertices \eqref{cubicV} take the form
\begin{multline}
{\cal A}^{\text{AdS}}_{\Delta_1 \Delta_2 |\Delta| \Delta_3 \Delta_4}\left(k_1,k_2;p_1;p_2;k_3,k_4\right)=  \int^{+i\infty}_{-i\infty}[{\rm d}s]_4[{\rm d}u]_2 \\ \times
{\cal A}^{\text{AdS}}_{\Delta_1 \Delta_2 |\Delta| \Delta_3 \Delta_4}\left(k_1,s_1;k_2,s_2;p_1,u_1;p_2,u_2;k_3,s_3;k_4,s_4\right),
\end{multline}
with
\begin{multline}\label{MBexch}
{\cal A}^{\text{AdS}}_{\Delta_1 \Delta_2 |\Delta| \Delta_3 \Delta_4}\left(k_1,s_1;k_2,s_2;p_1,u_1;p_2,u_2;k_3,s_3;k_4,s_4\right)= \frac{\Gamma\left(1+i\nu\right)\Gamma\left(1-i\nu\right)}{\pi} \csc\left(\pi \left(u_1+u_2\right)\right) \\ \times \omega_{\Delta}\left(u_1,u_2\right) {\cal A}^{\text{AdS}}_{\Delta_1 \Delta_2 \Delta}\left(s_1,k_1;s_2,k_2; u_1,p_1\right){\cal A}^{\text{AdS}}_{\Delta_3 \Delta_4 d-\Delta}\left(s_3,k_3;s_4,k_4; u_2,p_2\right),
\end{multline}
in terms of the constituent three-point contact diagrams \eqref{MBcontact} in Mellin space.

\paragraph{Identities.} Like for the propagators, Mellin space makes manifest certain identities satisfied by boundary correlation functions, such as those involving rotations by a phase. 

\vskip 4pt
One example is the relationship between EAdS boundary correlators and their counterparts in dS in the Bunch-Davies vacuum, which is manifest in Mellin space \cite{Sleight:2019hfp,Sleight:2020obc,Sleight:2021plv}. For example, considering the dS contact diagram generated by the vertex \eqref{npointvertex}, in the Bunch-Davies vacuum the contribution from the $\pm$ branch of the in-in contour is
\begin{align}\nonumber
\hspace*{-0.25cm}{}^{(0,\pm)}{\cal A}^{{\cal V}_{12 \ldots n}}_{\Delta_1 \Delta_2 \ldots \Delta_n}\left(s_1,k_1;s_2,k_2; \ldots ; s_n,k_n\right)&=\pm i \int_{-\infty}^0 \frac{{\rm d}\left(-\eta\right)}{\left(-\eta\right)^{d+1}}\,\left(-\eta\right)^{-\sum\limits^n_{i=1}\left(2s_i-\tfrac{d}{2}\right)}e^{\mp \left(s+\tfrac{1}{2}\left(\Delta_i-\tfrac{d}{2}\right)\right)\pi i}\\ & \hspace*{-5.5cm} \times \prod^n_{i=1}\,\frac{\Gamma\left(\frac{d}{2}-\Delta_i\right)}{4\pi}\Gamma\left(s_i+\tfrac{1}{2}\left(\tfrac{d}{2}-\Delta_i\right)\right)\Gamma\left(s_i-\tfrac{1}{2}\left(\tfrac{d}{2}-\Delta_i\right)\right) \left(\frac{k_i}{2}\right)^{-2s+\Delta-\tfrac{d}{2}}, \end{align}
which, using the Dirac delta function \eqref{MBdiracdelta}, immediately recovers \eqref{pmcontads}:
\begin{multline}
{}^{(0,\pm)}{\cal A}^{{\cal V}_{12 \ldots n}}_{\Delta_1 \Delta_2 \ldots \Delta_n}\left(k_1,k_2, \ldots ,k_n\right)= \pm i\, e^{\mp \frac{i \pi}{2}\left(\frac{(n-2)d}{2}+ i (\nu_1+\ldots+\nu_n)\right)}\left(\prod\limits^n _{i=1} c^{\text{dS-AdS}}_{\Delta_i}\left(-\eta_0\right)^{\Delta_i}\right)\\ \times {\cal A}^{\text{AdS}}_{\Delta_1 \ldots \Delta_n}\left(k_1,\ldots,k_n\right).
\end{multline}

\vskip 4pt
Other examples that we make use of in this work involve rotations of the boundary momenta. For example, for contact diagrams \eqref{MBcontact} we have 
\begin{multline}
{\cal A}^{\text{AdS}}_{\Delta_1 \Delta_2 \ldots \Delta_n}\left(s_1,e^{\pm \pi i}k_1;s_2,e^{\pm \pi i}k_2; \ldots ; s_m,e^{\pm \pi i}k_m; s_{m+1},k_{m+1}; \ldots s_n,k_n\right)\\
= e^{\pm \left(\Delta_1 + \ldots +\Delta_n-(n-1) d\right) \pi i}\\ \times {\cal A}^{\text{AdS}}_{\Delta_1 \Delta_2 \ldots \Delta_n}\left(s_1,k_1;s_2,k_2; \ldots ; s_m,k_m; s_{m+1},e^{\mp \pi i}k_{m+1}; \ldots s_n,e^{\mp \pi i}k_n\right),
\end{multline}
which also comes from applying the Dirac delta function in the Mellin variables. For $n=3$ this reduces to \eqref{3ptmomflip}. For exchange diagrams \eqref{MBexch}, a corollary of the above identity is that rotations of the internal momentum can be traded for rotations of the external momenta:
\begin{multline}
{\cal A}^{\text{AdS}}_{\Delta_1 \Delta_2 |\Delta| \Delta_3 \Delta_4}\left(k_1,k_2;{\bar p}^\pm_1;{\bar p}^{{\hat \pm}}_2;k_3,k_4\right) = \left(e^{\pm i \pi}\right)^{\tfrac{d}{2}-i\nu-i\nu_1-i\nu_2}\left(e^{{\hat \pm} i \pi}\right)^{\tfrac{d}{2}+i\nu-i\nu_3-i\nu_4}\\ \times {\cal A}^{\text{AdS}}_{\Delta_1 \Delta_2 |\Delta| \Delta_3 \Delta_4}\left({\bar k}^\mp_1,{\bar k}^\mp_2;p_1;p_2;{\bar k}^{\hat \mp}_3,{\bar k}^{\hat \mp}_4\right).
\end{multline}

\subsection{Extension to $\alpha$-vacua}
\label{subsec::MSextalpha}

In \cite{Sleight:2019mgd,Sleight:2019hfp,Sleight:2020obc,Sleight:2021plv} the Mellin space representation of boundary correlators in (EA)dS was introduced assuming the absence of singularities in folded configurations of the boundary momenta. This is sufficient for the study of boundary correlators in AdS and the Bunch-Davies vacuum in dS but, as explained in section \ref{sec::CWI}, for a generic dS invariant vacuum this assumption must be relaxed.

\vskip 4pt
In the following section we extend the Mellin space representation of boundary correlation functions to allow for singularities in folded momentum configurations. This is first carried out from a bulk perspective in section \ref{subsubsec::MSbulk}, i.e. at the level of the Feynman rules (\eqref{ininbubu} and \eqref{ininbubo}) for late-time correlators in $\alpha$ vacua, and then from a boundary perspective in section \ref{subsec::ABP} at the level of the conformal ward identities (building upon section 3.1 of \cite{Sleight:2021plv}).

\subsubsection{Bulk perspective}
\label{subsubsec::MSbulk}

In section \ref{subsec::alphafromBD} we saw that the propagators for late-time correlators in a generic $\alpha$ vacuum could be re-written in terms of the corresponding propagators in the Bunch-Davies vacuum $\left(\alpha=0\right)$ with appropriate rotatations of the modulus of boundary momenta. Since in Mellin space the dependence on the modulus of the boundary momentum enters as a power, with the exponent given by the corresponding Mellin variable (see \eqref{buboMB}), the Mellin space representation of propagators in a generic $\alpha$ vacuum are given by those in the Bunch-Davies vacuum appropriately dressed by phases in the Mellin variables. 

\vskip 4pt
In particular, from the identities \eqref{bubuan} and \eqref{buboan} it follows immediately that propagators in a generic $\alpha$ vacuum take the following form in Mellin space:
  \begin{align}\nonumber
G^{(\alpha)}_{++}(u_1,k;u_2,k)&=\left[P^+_{\Delta_+}+e^{-2\nu\pi }P^-_{\Delta_+}\, e^{2\left(u_1+u_2\right)\pi i} \right]G^{(0)}_{\Delta_+,++}(u_1,k;u_2,k) + (\Delta_+ \to \Delta_-),\\ \nonumber
G^{(\alpha)}_{--}(u_1,k;u_2,k)&=\left[M^-_{\Delta_+} + \, e^{2\nu \pi }  M^+_{\Delta_+}\, e^{-2\left(u_1+u_2\right)\pi i}\right]G^{(0)}_{\Delta_+,--}(u_1,k;u_2,k)+ (\Delta_+ \to \Delta_-), \\ \nonumber
G^{(\alpha)}_{-+}(u_1,k;u_2,k)& =  \left[\cosh^2\alpha \, +\sinh^2\alpha \,e^{-2\left(u_1-u_2\right)\pi i} \right. \\  \nonumber & \hspace*{1cm}\left.-\frac{1}{2}\sinh 2\alpha\,e^{+\nu\pi} \left( e^{i\beta}\, e^{2u_2\pi i} + e^{-i\beta} \, e^{-2 u_1\pi i} \right)\right]G^{(0)}_{\Delta_+,-+}(u_1,k;u_2,k)\\ &+ (\Delta_+ \to \Delta_-), \nonumber \\ \nonumber
G^{(\alpha)}_{+-}(u_1,k;u_2,k) &= \left[ \cosh^2\alpha \, +\sinh^2\alpha \, e^{2\left(u_1-u_2\right)\pi i}\right.\\ \nonumber
 &\hspace*{1cm}\left.-\frac{1}{2}\sinh 2\alpha\,e^{-\nu\pi}( e^{i\beta}\, e^{-2u_2\pi i} + e^{-i\beta} \, e^{2u_1\pi i})\right]G^{(0)}_{\Delta_+,+-}(u_1,k;u_2,k)\\ &+ (\Delta_+ \to \Delta_-), \label{bubufromBDa}
\end{align}  
and
\begin{subequations}
 \label{bubofromBDa}
\begin{align}
K^{(\alpha)}_{\Delta,\, +}\left(s,k\right) &= \left[P^+_{\Delta} + P^-_{\Delta}  e^{-\nu \pi} e^{2s\pi i} \right]K^{(0)}_{\Delta,\, +}\left(s,k\right),\\
K^{(\alpha)}_{\Delta,\, -}\left(s,k\right) &= \left[ M^-_{\Delta} + M^+_{\Delta} e^{\nu \pi } e^{-2s\pi i} \right] K^{(0)}_{\Delta,\, -}\left(s,k\right).
\end{align}   
\end{subequations}

\vskip 4pt
The upshot is that in Mellin space perturbative late-time correlators in a generic $\alpha$ vacuum are given by the Mellin space representation of the corresponding process in the Bunch-Davies vacuum dressed by phases in the Mellin variables. Owing to the Dirac delta function \eqref{DDMellin} enforcing invariance under dilatations, such phases moreover can be expressed purely in terms of Mellin variables associated to external legs.

\vskip 4pt 
For example, in Mellin space the three-point contact diagram \eqref{3ptcontactdS} in a generic $\alpha$ vacuum is given by 
\begin{multline}\label{MS3pt}
    {}^{(\alpha)}{\cal A}^{{\cal V}_{123}}_{\Delta_1 \Delta_2 \Delta_3}\left(s_1,k_1;s_2,k_2;s_3,k_3\right) =  i\left(\prod\limits^3 _{i=1} c^{\text{dS-AdS}}_{\Delta_i}\left(-\eta_0\right)^{\Delta_i}\right)\sum_{\pm}e^{\mp \frac{i \pi}{4}(d+2 i (\nu_1+\nu_2+\nu_3))}\\ \times \left\{  {}^{\left(0\right)}C^\pm_{\Delta_1\Delta_2\Delta_3} + \sinh (2 \alpha )  {}^{\left(\alpha\right)}C^\pm_{\Delta_1\Delta_2\Delta_3}
     \left(  e^{\pm 2s_1 \pi i}+e^{\pm 2s_2 \pi i} +e^{\pm 2s_3 \pi i}\right)\right\}\\ \times {\cal A}^{\text{AdS}}_{\Delta_1 \Delta_2 \Delta_3}\left(s_1,k_1;s_2,k_2;s_3,k_3\right).
\end{multline}

\subsubsection{Boundary perspective}
\label{subsec::ABP}

The Mellin space structure of late-time correlators in a generic de Sitter invariant vacuum can also be understood from a boundary perspective, following (and building on) section 3.1 of \cite{Sleight:2021plv}. The (EA)dS isometry transformations act on the boundary as conformal transformations and boundary correlators in these spaces are therefore constrained by conformal ward identities. In the following we shall review the implications of such conformal ward identities in Mellin space, focusing for simplicity on three-point functions.

\vskip 4pt
In addition to translations and rotations, the conformal group is generated by dilatations and special conformal transformations, which in momentum space read, respectively,
\begin{subequations}
\begin{align}
{\cal D} &= -\left(\Delta-d\right)+k \partial_k,\\
{\cal K}_{k_i} &= 2\left(\Delta-d\right)\partial_{k_i}-2k^j \partial_{k_j} \partial_{k_i}+k^i \partial^2_{k_j}.
\end{align}
\end{subequations}
Translations and rotations require, in the usual way, that three-point functions are proportional to a momentum conserving delta function with coefficient given by a function of the magnitudes $k_i=|\vec{k}_i|$ of the boundary momenta: 
\begin{equation}\label{momcons}
 F_{\Delta_1\Delta_2\Delta_3}\left(\vec{k}_1,\vec{k}_2,\vec{k}_3\right)=\left(2\pi\right)^d \delta^{\left(d\right)}\left(\vec{k}_1+\vec{k}_2+\vec{k}_3\right)F^\prime_{\Delta_1\Delta_2\Delta_3}\left(k_1,k_2,k_3\right),
\end{equation}
where $F_{\Delta_1\Delta_2\Delta_3}$ is a momentum space three-point function of operators with scaling dimension $\Delta_i$.

\vskip 4pt
To solve the Ward identities associated to dilatations and special conformal transformations we go to Mellin space, where
\begin{equation}
F^\prime_{\Delta_1\Delta_2\Delta_3}\left(k_1,k_2,k_3\right) = \int^{+i\infty}_{-i\infty}[{\rm d}s]_3\,F_{\Delta_1\Delta_2\Delta_3}\left(s_1,s_2,s_3\right) \prod^3_{i=1}\left(\frac{k_i}{2}\right)^{-2s_i+i\nu_i},
\end{equation}
and $F_{\Delta_1\Delta_2\Delta_3}\left(s_1,s_2,s_3\right)$ is the Mellin transform of $F^\prime_{\Delta_1\Delta_2\Delta_3}\left(k_1,k_2,k_3\right)$ with respect to the $k_i$.
The dilatation Ward identity is given by
\begin{equation}\label{Dward}
0=\left(-d+\sum\limits^3_{i=1}{\cal D}_j\right)F^\prime_{\Delta_1\Delta_2\Delta_3}\left(k_1,k_2,k_3\right),
\end{equation}
which in Mellin space translates into
\begin{equation}
0=\int^{+i\infty}_{-i\infty}[{\rm d}s]_3\,\left(\tfrac{d}{2}-2\left(s_1+s_2+s_3\right)\right)F_{\Delta_1\Delta_2\Delta_3}\left(s_1,s_2,s_3\right) \prod^3_{i=1}\left(\frac{k_i}{2}\right)^{-2s_i+i\nu_i}.
\end{equation}
This implies that $F_{\Delta_1\Delta_2\Delta_3}\left(s_1,s_2,s_3\right)$ takes the form:
\begin{equation}
F_{\Delta_1\Delta_2\Delta_3}\left(s_1,s_2,s_3\right) = 2\pi i\, \delta\left(\tfrac{d}{4}-s_1-s_2-s_3\right)F^\prime_{\Delta_1\Delta_2\Delta_3}\left(s_1,s_2,s_3\right),
\end{equation}
which is the Mellin space analogue of momentum conservation \eqref{momcons} implied by translation invariance.

\vskip 4pt
The Ward identity associated to special conformal transformations,
\begin{equation}
\left(\prod\limits^3_{i=1}{\cal K}^j_{k_i}\right)F^\prime_{\Delta_1\Delta_2\Delta_3}\left(k_1,k_2,k_3\right)=0.
\end{equation}
In \cite{Bzowski:2013sza} this was solved by reducing the Ward identity to two independent scalar equations:
\begin{subequations}\label{SKWI}
\begin{align}
    0&=\left[\left(\frac{\partial^2}{\partial k^2_1}+\frac{d+1-2\Delta_1}{k_1}\frac{\partial}{\partial k_1}\right)-\left(\frac{\partial^2}{\partial k^2_3}+\frac{d+1-2\Delta_3}{k_3}\frac{\partial}{\partial k_3}\right)\right]F^\prime_{\Delta_1\,\Delta_2\,\Delta_3}\left(k_1,k_2,k_3\right),\\
    0&=\left[\left(\frac{\partial^2}{\partial k^2_2}+\frac{d+1-2\Delta_2}{k_2}\frac{\partial}{\partial k_2}\right)-\left(\frac{\partial^2}{\partial k^2_3}+\frac{d+1-2\Delta_3}{k_3}\frac{\partial}{\partial k_3}\right)\right]F^\prime_{\Delta_1\,\Delta_2\,\Delta_3}\left(k_1,k_2,k_3\right),
\end{align}
\end{subequations}
 which is achieved by taking $\vec{k}_1$ and $\vec{k}_2$ to be independent momenta. In Mellin space these translate into the following system of difference relations:
\begin{multline}
\left(s_1-1+\tfrac{1}{2}\left(\tfrac{d}{2}-\Delta_1\right)\right)\left(s_1-1-\tfrac{1}{2}\left(\tfrac{d}{2}-\Delta_1\right)\right)F^\prime_{\Delta_1\Delta_2\Delta_3}\left(s_1-1,s_2,s_3\right)\\
=\left(s_3-1+\tfrac{1}{2}\left(\tfrac{d}{2}-\Delta_3\right)\right)\left(s_3-1-\tfrac{1}{2}\left(\tfrac{d}{2}-\Delta_3\right)\right)F^\prime_{\Delta_1\Delta_2\Delta_3}\left(s_1,s_2,s_3-1\right),
\end{multline}
\begin{multline}
\left(s_2-1+\tfrac{1}{2}\left(\tfrac{d}{2}-\Delta_2\right)\right)\left(s_2-1-\tfrac{1}{2}\left(\tfrac{d}{2}-\Delta_2\right)\right)F^\prime_{\Delta_1\Delta_2\Delta_3}\left(s_1,s_2-1,s_3\right)\\
=\left(s_3-1+\tfrac{1}{2}\left(\tfrac{d}{2}-\Delta_3\right)\right)\left(s_3-1-\tfrac{1}{2}\left(\tfrac{d}{2}-\Delta_3\right)\right)F^\prime_{\Delta_1\Delta_2\Delta_3}\left(s_1,s_2,s_3-1\right),
\end{multline}
which are solved by \cite{Sleight:2021plv}
\begin{multline}
F^\prime_{\Delta_1\Delta_2\Delta_3}\left(s_1,s_2,s_3\right) = p_{\Delta_1\Delta_2\Delta_3}\left(s_1,s_2,s_3\right) \\ \times \prod^3_{i=1} \Gamma\left(s_i+\tfrac{1}{2}\left(\tfrac{d}{2}-\Delta_i\right)\right)\Gamma\left(s_i-\tfrac{1}{2}\left(\tfrac{d}{2}-\Delta_i\right)\right),
\end{multline}
where $p_{\Delta_1\Delta_2\Delta_3}\left(s_1,s_2,s_3\right)$ is a periodic function of unit period in the Mellin variables $s_1$, $s_2$ and $s_3$, which is furthermore constrained by the requirement that the Mellin integrals converge. Different choices for  $p_{\Delta_1\Delta_2\Delta_3}\left(s_1,s_2,s_3\right)$ yield different solutions, which are four in total since we seek solutions to two second order equations \eqref{SKWI}.

\vskip 4pt 
The solution with no folded singularities is $p_{\Delta_1\Delta_2\Delta_3}\left(s_1,s_2,s_3\right)=\text{constant}$, which corresponds to the conformal structure accounting for boundary correlators in EAdS and in the Bunch-Davies vacuum of dS. This is the Mellin space representation of the ``triple-$K$" integral solution \cite{Bzowski:2013sza} to the conformal Ward identities, given by \eqref{nK} with $n=3$. It is therefore manifest that in Mellin space the other solutions to the conformal ward identities are given by dressings of this solution by phases of unit period in the Mellin variables, as in the expression \eqref{MS3pt} for late-time three-point functions in $\alpha$ vacua.

\vskip 4pt
The four solutions to the conformal Ward identities can all be written in the ``triple Bessel" form (equation (3.23) of \cite{Bzowski:2013sza}):
\begin{equation}
k^{\Delta_1-\tfrac{d}{2}}_1k^{\Delta_2-\tfrac{d}{2}}_2k^{\Delta_3-\tfrac{d}{2}}_3\int^\infty_0 {\rm d}z\,z^{\tfrac{d}{2}-1} I_{\pm(\Delta_1-\frac{d}{2})}\left(k_1z\right)I_{{\hat \pm}(\Delta_2-\frac{d}{2})}\left(k_2 z\right)K_{\Delta_3-\frac{d}{2}}\left(k_3 z\right),
\end{equation}
in terms of modified Bessel functions of both the first and second kinds. In Mellin space, these correspond to applications of the projectors \eqref{projectors} to the ``triple-$K$" solution above: 
\begin{equation}
p_{\Delta_1\Delta_2\Delta_3}\left(s_1,s_2,s_3\right) \propto e^{-s_1 \pi  i } \sin \left(\pi  \left(s_1\mp \tfrac{i \nu_1}{2}\right)\right)e^{-s_2 \pi  i } \sin \left(\pi  \left(s_2\,{\hat \mp}\, \tfrac{i \nu_2}{2}\right)\right).
\end{equation}
In particular, using Cauchy's residue theorem it is straightforward to show that 
\begin{multline}
\left(\frac{k}{2}\right)^{+i\nu} z^{\tfrac{d}{2}} I_{i\nu}\left(kz\right) = \int^{+i\infty}_{-i\infty}\frac{{\rm d}s}{2\pi i} e^{-\left(s-\frac{i \nu }{2}\right)\pi  i} \sin \left(\pi  \left(s-\tfrac{i \nu }{2}\right)\right) \\ \times \Gamma(s+\tfrac{i\nu}{2})\Gamma(s-\tfrac{i\nu}{2})\left(\frac{k}{2}\right)^{-2s+i\nu}.
\end{multline}

\subsection{Regularisation}
\label{app::MBreg}

For Mellin-Barnes integrals the contour of integration runs parallel to the imaginary axis and is suitably indented so that it separates sequences of poles of the type $\Gamma\left(s_i+a_i\right)$ from those of the type $\Gamma\left(-s_i+b_i\right)$. This contour prescription is well defined for parameters $a_i$ and $b_i$ such that poles from $\Gamma$-functions of the type $\Gamma\left(s_i+a_i\right)$ do not collide poles from those of the type $\Gamma\left(-s_i+b_i\right)$. This is always the case for Principal Series representations where $\Delta_i = \frac{d}{2}+i\nu_i$, $\nu_i \in \mathbb{R}$, and such poles are therefore always separated. This is not always the case for the other unitary representations (i.e. the complementary series representations, see section \ref{sec::SFEPP}), where $\nu_i$ is imaginary, which can lead to pinching of the integration contour. The latter gives rise to singularities which require careful regularisation and in some cases also renormalisation.\footnote{See \cite{Bzowski:2013sza,Bzowski:2015pba,Bzowski:2015yxv,Bzowski:2017poo,Bzowski:2018fql,Bzowski:2023nef} for authoritative works on regularisation and renormalisation of momentum space correlation functions in CFT.} 

\vskip 4pt
An example that we encounter in this work is the three-point contact diagram \eqref{3pt} with $\nu_i = \frac{3i}{2}$, which in $d=3$ correspond to massless scalar fields. Taking for the moment $d$ arbitrary, from the Mellin space expression \eqref{MBcontact} we have (following section 3.4 of \cite{Sleight:2019mgd})
\begin{multline}
\hspace*{-0.5cm} {\cal A}^{\text{AdS}}_{000}\left(k_1,k_2, k_3\right)=-\frac{2}{k^3_2k^3_3} \int^{+i\infty}_{-i\infty}\frac{{\rm d}s_1}{2\pi i}\left(d-4 (s_1+1)\right) \left(2 s_1-\tfrac{1}{2}\right) \Gamma \left(2 s_1-\tfrac{3}{2}\right) \Gamma \left(\tfrac{d}{2}-2 s_1-3\right) \\ \times  \left(\frac{1}{2} k_2 k_3 (d-4 s_1-6)+(k_2+k_3)^2\right)\left(k_2+k_3\right)^{-\frac{d}{2}+2 s_1+1}k_1^{-2 s_1-\frac{3}{2}},
\end{multline}
where we eliminated the $s_3$ integral using the Dirac delta function \eqref{MBdiracdelta} and the $s_2$ integral by applying Cauchy's residue theorem. For the remaining $s_1$ integral, the $\Gamma$-function poles are
\begin{align}
s_1 & = \frac{3}{4}-\frac{n}{2}, \qquad n = 0, 1, 2, 3, \ldots \\
s_1 & = \frac{d-6}{4}+\frac{m}{2}, \qquad m = 0, 1, 2, 3, \ldots\,,
\end{align}
which are overlapping for $d \leq 9$. Keeping $d$ arbitrary, evaluating the $s_1$ integral by closing the contour to the right one obtains
\begin{multline}
    {\cal A}^{\text{AdS}}_{000}\left(k_1,k_2,k_3\right) = \frac{1}{3 k^3_1k^3_2k^3_3}\left[2 k^2_1 (k_2+k_3)+2 k_1 \left(k_2^2-k_2 k_3+k_3^2\right)+2 k_2 k_3 (k_2+k_3)\right. \\ \left. +\frac{4 \left(k_1^3+k_2^3+k_3^3\right)}{d-3}-\frac{1}{3} (6 \gamma -11) \left(k_1^3+k_2^3+k_3^3\right) \right. \\ \left. - \left(k_1^3+k_2^3+k_3^3\right) (2 \log (k_1+k_2+k_3)+1)\right].
\end{multline}
Note that there is a simple pole for $d=3$ which requires renormalisation. This divergence is local and therefore, in the corresponding dS contact diagram \eqref{massless3pt}, can be cancelled in the in-in formalism by adding local counterterms at the future boundary of dS \cite{Bzowski:2023nef}.

\end{appendix}
\clearpage
\bibliographystyle{JHEP}
\bibliography{refs}

\end{document}